\documentstyle[12pt,epsfig]{article}
 
\newcommand{\agt}{\,\rlap{\lower 3.5 pt \hbox{$\mathchar \sim$}} \raise 1pt
 \hbox {$>$}\,}
\newcommand{\alt}{\,\rlap{\lower 3.5 pt \hbox{$\mathchar \sim$}} \raise 1pt
 \hbox {$<$}\,}

\catcode`@=11
\newcount\@tempcntc
\def\@citex[#1]#2{\if@filesw\immediate\write\@auxout{\string\citation{#2}}\fi
  \@tempcnta\z@\@tempcntb\m@ne\def\@citea{}\@cite{\@for\@citeb:=#2\do
    {\@ifundefined
       {b@\@citeb}{\@citeo\@tempcntb\m@ne\@citea\def\@citea{,}{\bf ?}\@warning
       {Citation `\@citeb' on page \thepage \space undefined}}%
    {\setbox\z@\hbox{\global\@tempcntc0\csname b@\@citeb\endcsname\relax}%
     \ifnum\@tempcntc=\z@ \@citeo\@tempcntb\m@ne
       \@citea\def\@citea{,}\hbox{\csname b@\@citeb\endcsname}%
     \else
      \advance\@tempcntb\@ne
      \ifnum\@tempcntb=\@tempcntc
      \else\advance\@tempcntb\m@ne\@citeo
      \@tempcnta\@tempcntc\@tempcntb\@tempcntc\fi\fi}}\@citeo}{#1}}
\def\@citeo{\ifnum\@tempcnta>\@tempcntb\else\@citea\def\@citea{,}%
  \ifnum\@tempcnta=\@tempcntb\the\@tempcnta\else
   {\advance\@tempcnta\@ne\ifnum\@tempcnta=\@tempcntb \else \def\@citea{--}\fi
    \advance\@tempcnta\m@ne\the\@tempcnta\@citea\the\@tempcntb}\fi\fi}
\catcode`@=12

\begin{document}

\title{\vskip-3cm{\baselineskip14pt
\centerline{\normalsize DESY 00-158\hfill ISSN 0418-9833}
\centerline{\normalsize MPI/PhT/2000-33\hfill}
\centerline{\normalsize hep-ph/0011155\hfill}
\centerline{\normalsize November 2000\hfill}
}
\vskip1.5cm
Testing the Universality of Fragmentation Functions
}
\author{{\sc B.A. Kniehl,$^1$ G. Kramer,$^1$ B. P\"otter$^2$}\\
{\normalsize $^1$ II. Institut f\"ur Theoretische Physik, Universit\"at
Hamburg,}\\
{\normalsize Luruper Chaussee 149, 22761 Hamburg, Germany}\\
{\normalsize $^2$ Max-Planck-Institut f\"ur Physik
(Werner-Heisenberg-Institut),}\\
{\normalsize F\"ohringer Ring 6, 80805 Munich, Germany}}

\date{}

\maketitle

\thispagestyle{empty}

\begin{abstract}
Using fragmentation functions for charged pions, charged kaons, and 
(anti)pro\-tons recently extracted from experimental data of $e^+e^-$
annihilation at the $Z$-boson resonance and at centre-of-mass energy
$\sqrt s=29$~GeV, we perform a global study of inclusive charged-hadron 
production in $p\bar p$, $\gamma p$, and $\gamma\gamma$ collisions at 
next-to-leading order in the parton model of quantum chromodynamics.
Comparisons of our results with $p\bar p$ data from CERN S$p\bar p$S and the
Fermilab Tevatron, $\gamma p$ data from DESY HERA, and $\gamma\gamma$ data 
from CERN LEP2 allow us to test the universality of the fragmentation
functions predicted by the factorization theorem.
Furthermore, we perform comparisons with $e^+e^-$-annihilation data from LEP2
so as to test the scaling violations predicted by the Altarelli-Parisi
evolution equations.
\medskip

\end{abstract}

\newpage

\section{Introduction}

In the framework of the QCD-improved parton model, the inclusive production of
single hadrons is described by means of fragmentation functions (FFs),
$D_a^h(x,\mu^2)$.
The value of $D_a^h(x,\mu^2)$ corresponds to the probability for the parton
$a$ produced at short distance $1/\mu$ to form a jet that includes the hadron
$h$ carrying the fraction $x$ of the longitudinal momentum of $a$.
Unfortunately, it is not yet possible to derive the FFs from first principles,
in particular for hadrons with masses smaller than or comparable to the
asymptotic scale parameter, $\Lambda$.
However, given their $x$ dependence at some energy scale $\mu$, the evolution
with $\mu$ may be computed perturbatively in QCD using the timelike 
Altarelli-Parisi equations \cite{gri}.
This allows us to test QCD quantitatively within one experiment observing
single hadrons at different values of centre-of-mass (CM) energy $\sqrt s$ (in
the case of $e^+e^-$ annihilation) or transverse momentum $p_T$ (in the case
of scattering).
Moreover, the factorization theorem guarantees that the $D_a^h(x,\mu^2)$
functions are independent of the process in which they have been determined,
and represent a universal property of $h$.
This enables us to make quantitative predictions for other types of
experiments as well.

During the last five years, the experiments at the CERN Large
Electron-Positron Collider (LEP1) and the SLAC Linear Collider (SLC) have
provided us with a wealth of high-precision information on how partons
fragment into low-mass charged hadrons.
The data partly comes as light-, $c$-, and $b$-quark-enriched samples without
\cite{Al,A,D,O} or with identified final-state hadrons ($\pi^\pm$, $K^\pm$,
and $p/\bar p$) \cite{A1,D1,S} or as gluon-tagged three-jet samples without
hadron identification \cite{Ag,Dg}.
This new situation motivated us to update, refine, and extend a former
analysis performed by two of us together with Binnewies \cite{bkk0,bkk} by
generating new leading-order (LO) and next-to-leading-order (NLO) sets of
$\pi^\pm$, $K^\pm$, and $p/\bar p$ FFs \cite{kkp}.\footnote{%
A FORTRAN subroutine which returns the values of the $D_a^h(x,\mu^2)$
functions for given values of $x$ and $\mu^2$ may be downloaded from the URL
{\tt http://www.desy.de/\~{}poetter/kkp.html} or obtained upon request from
the authors.}
By also including in our fits $\pi^\pm$, $K^\pm$, and $p/\bar p$ data (without
flavour separation) from SLAC PEP \cite{T}, with CM energy $\sqrt s=29$~GeV,
we obtained a handle on the scaling violations in the FFs, which enabled us to
determine the strong-coupling constant.
We found $\alpha_s^{(5)}(M_Z)=0.1181\pm0.0085$ at LO and
$\alpha_s^{(5)}(M_Z)=0.1170\pm0.0073$ at NLO \cite{kkp1}.
These results are in perfect agreement with what the Particle Data Group 
currently quotes as the world average, $\alpha_s^{(5)}(M_Z)=0.1181\pm0.002$
\cite{pdg}.
Our strategy was to only include in our fits LEP1 and SLC data with both
flavour separation and hadron identification \cite{A1,D1,S},\footnote{%
Contrary to what is stated in the respective figure captions, the squares
pertaining to the upmost curves shown in Figs.~2--4 of Ref.~\cite{kkp} refer
to OPAL data \cite{O}.}
gluon-tagged three-jet samples with a fixed gluon-jet energy \cite{Ag}, and
the $\pi^\pm$, $K^\pm$, and $p/\bar p$ data sets from the pre-LEP1/SLC era
with the highest statistics and the finest binning in $x$ \cite{T}.
Other data served us for cross checks.
In particular, we probed the scaling violations in the FFs through comparisons
with $\pi^\pm$, $K^\pm$, and $p/\bar p$ data from DESY DORIS and PETRA, with
CM energies between 5.4 and 34~GeV \cite{low}.
Furthermore, we tested the gluon FF, which enters the unpolarized cross
section only at NLO, by comparing our predictions for the longitudinal cross
section, where it already enters at LO, with available data \cite{Al,Dl}.
Finally, we directly compared our gluon FF with the one recently measured by
DELPHI in three-jet production with gluon identification as a function of $x$
at various energy scales $\mu$ \cite{Dg}.
All these comparisons led to rather encouraging results.
We also verified that our FFs satisfy reasonably well the momentum sum
rules, which we did not impose as constraints on our fits.

The purpose of this paper is twofold.
On the one hand, we wish to extend our previous tests of scaling violations 
\cite{kkp} to higher energy scales by confronting new data on inclusive 
charged-hadron production in $e^+e^-$ annihilation from LEP2 \cite{D2,O2},
with $\sqrt s$ ranging from 133~GeV up to 189~GeV, with NLO predictions based
on our FFs.
On the other hand, we wish to quantitatively check the universality of our FFs
by making comparisons with essentially all available high-statistics data on
inclusive charged-hadron production in colliding-beam experiments.
This includes $p\bar p$ data from the UA1 \cite{UA1a,UA1b,UA1c} and UA2
\cite{UA2a,UA2b} Collaborations at the CERN Super Proton-Antiprotron
Synchrotron (S$p\bar p$S) and from the CDF Collaboration \cite{CDF} at the
Fermilab Tevatron, $\gamma p$ data from the H1 \cite{H1} and ZEUS \cite{ZEUS}
Collaborations at DESY HERA, and $\gamma\gamma$ data from the OPAL 
Collaboration \cite{Ogg} at LEP2.

Very recently, alternative sets of NLO FFs for $\pi^\pm$, $K^\pm$ \cite{kre},
and charged hadrons \cite{kre,bou} have become available.
They are based on different collections of experimental data and on additional
theoretical assumptions.
In Ref.~\cite{kre}, power laws were assumed to implement a hierarchy among the
valence- and sea-quark FFs.
In Ref.~\cite{bou}, the renormalization and factorization scales were 
identified and adjusted according to the principle of minimal sensitivity
\cite{ste}.
In order to estimate the present systematic uncertainties in the FFs, it is
useful to compare NLO predictions for inclusive charged-hadron production
consistently evaluated with the three new-generation FF sets 
\cite{kkp,kre,bou}.

This work is organized as follows.
In Section~\ref{sec:two}, we present comparisons of our NLO predictions for
inclusive charged-hadron production with $e^+e^-$ data from LEP2 \cite{D2,O2},
$p\bar p$ data from S$p\bar p$S \cite{UA1a,UA1b,UA1c,UA2a,UA2b} and the
Tevatron \cite{CDF}, $\gamma p$ data from HERA \cite{H1,ZEUS}, and
$\gamma\gamma$ data from LEP2 \cite{Ogg}.
In Section~\ref{sec:three}, we compare our NLO FFs with those of 
Refs.~\cite{kre,bou} by studying inclusive charged-hadron production in 
$e^+e^-$ annihilation at various CM energies for the case of quark-flavour
separation.
Our conclusions are summarized in Section~\ref{sec:four}.
In the Appendix, we list convenient parameterizations of the $x$ and $\mu^2$ 
dependences of our $\pi^\pm$, $K^\pm$, and $p/\bar p$ FFs at LO and NLO
\cite{kkp}.

\section{Global Analysis of Collider Data
\label{sec:two}}

Throughout this paper, we work at NLO in the $\overline{\mathrm{MS}}$
renormalization and factorization scheme with $n_f=5$ active quark flavours.
We choose the renormalization scale $\mu$ and the factorization scale $M_f$ to
be $\mu=M_f=\xi\sqrt s$ in the case of $e^+e^-$ annihilation and
$\mu=M_f=\xi p_T$ otherwise, where the dimensionless parameter $\xi$ is
introduced to determine the theoretical uncertainty due to scale variations.
As usual, we allow for variations of $\xi$ between $1/2$ and 2 around the
default value 1.
As for the parton density functions (PDFs) of the proton, we employ set CTEQ5M
provided by the CTEQ Collaboration \cite{CTEQ5}.
As for the photon PDFs, we use the set by Aurenche, Fontannaz, and Guillet 
(AFG) \cite{AFG} as default and those by Gl\"uck, Reya, and Vogt (GRV)
\cite{GRV} and by Gordon and Storrow (GS) \cite{GS} for comparison.
As for the charged-hadron FFs, we take the sum of our $\pi^\pm$, $K^\pm$, and
$p\bar p$ FFs \cite{kkp} as default.
In Section~\ref{sec:three}, we also use the charged-hadron sets by Kretzer (K)
\cite{kre} and by Bourhis, Fontannaz, Guillet, and Werlen (BFGW) \cite{bou}
for comparison.
We evaluate $\alpha_s^{(n_f)}(\mu)$ from the two-loop formula \cite{pdg} with
the value of $\Lambda_{\overline{\mathrm{MS}}}^{(5)}$ that belongs to the 
respective FF set, namely
$\Lambda_{\overline{\mathrm{MS}}}^{(5)}=213$~MeV for Ref.~\cite{kkp},
$\Lambda_{\overline{\mathrm{MS}}}^{(5)}=168$~MeV for Ref.~\cite{kre}, and
$\Lambda_{\overline{\mathrm{MS}}}^{(5)}=215$~MeV for Ref.~\cite{bou}.
These values are in the same ball park as the one pertaining to the CTEQ5M
\cite{CTEQ5} PDFs, $\Lambda_{\overline{\mathrm{MS}}}^{(5)}=226$~MeV.
The considered photon PDFs \cite{AFG,GRV,GS} are all implemented with
$\Lambda_{\overline{\mathrm{MS}}}^{(4)}=200$~MeV, which corresponds to
$\Lambda_{\overline{\mathrm{MS}}}^{(5)}=137$~MeV for a $b$-quark mass of
$m_b=4.5~$GeV.
The NLO cross sections for the partonic subprocesses $e^+e^-\to c+X$,
$ab\to c+X$, $\gamma b\to c+X$, and $\gamma\gamma\to c+X$, where $a,b,c$ 
denote partons (quarks and gluons) and $X$ represents the residual final-state
partons, were found in Refs.~\cite{alt}, \cite{ave}, \cite{aur,gor}, and
\cite{gor,aur1}, respectively.
The NLO formalism for calculating the hadronic cross sections for $e^+e^-$
annihilation, hadroproduction in $p\bar p$ collisions, and photoproduction in
$e^\pm p$ and $e^+e^-$ collisions is described in 
Refs.~\cite{bkk,bor,bor1,bkk1}, respectively.
In the following, we always consider the sum of positively and negatively 
charged hadrons.
While the experimental data are sampled in bins, we present our theoretical 
predictions as continuous distributions.
This should be justified as long as the bins are narrow or the data is
corrected for finite bin widths.

\boldmath
\subsection{$e^+e^-$ Annihilation}
\unboldmath

\begin{figure}[ht]
\begin{center}
\centerline{\epsfig{figure=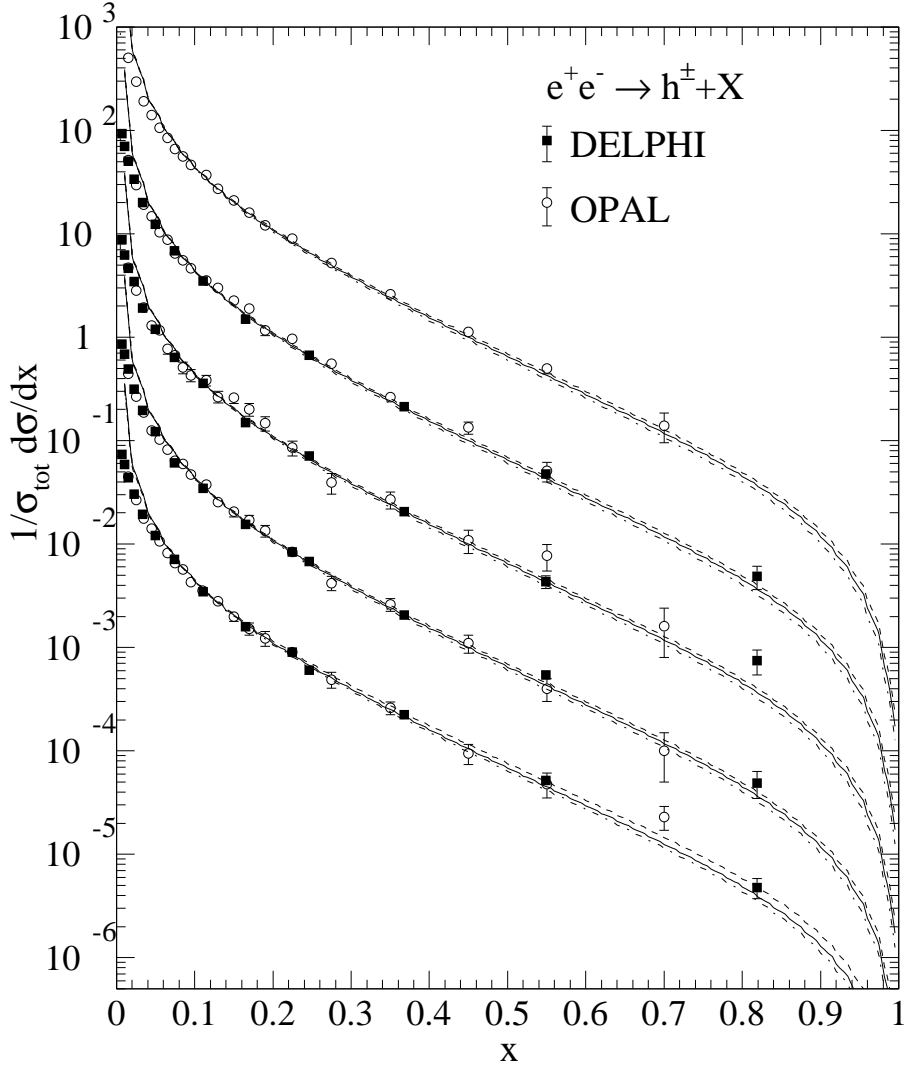,height=16cm}}
\caption{Normalized differential cross section
$(1/\sigma_{\mathrm{tot}})d\sigma/dx$ (in pb) of inclusive charged-hadron
production in $e^+e^-$ annihilation $e^+e^-\to h^\pm+X$ as a function of
scaled momentum $x$ at CM energies $\protect\sqrt s=133$, 161, 172, 183, and
189~GeV (from bottom to top in this order).
The NLO predictions for $\xi=1/2$ (dashed lines), 1 (solid lines), and 2 
(dot-dashed lines) are compared with data from DELPHI \protect\cite{D2} (solid
boxes) and OPAL \protect\cite{O2} (open circles).
Each set of curves is rescaled relative to the nearest upper one by a factor
of 1/10.}
\label{fig:do}
\end{center}
\end{figure}

In Ref.~\cite{kkp}, we already tested the scaling violations in our FFs,
predicted by the timelike $\mu^2$ evolution equations, by making comparisons 
with $e^+e^-$ data of inclusive $\pi^\pm$, $K^\pm$, and $p/\bar p$ production
in the energy range from $\sqrt s=5.2$~GeV to 44~GeV \cite{low}.
The advent of LEP2 data on $e^+e^-\to h^\pm+X$ with $\sqrt s=133$, 161, 172,
183, and 189~GeV, taken by DELPHI \cite{D2} and OPAL \cite{O2}, enables us to
gradually extend these tests towards high energies.
This data is given as the normalized differential cross section
$(1/\sigma_{\mathrm{tot}})d\sigma/dx$ as a function of the scaled momentum
$x=2p/\protect\sqrt s$, where $p$ is the absolute value of the three-momentum
of the final-state hadron.
In Fig.~\ref{fig:do}, it is compared with our NLO predictions for $\xi=1/2$, 
1, and 2.

We observe that the scale variation of the latter is rather small, about
$\pm1\%$ at $x=0.1$ and $\pm10\%$ at $x=0.8$, although compensating effects on
the renormalization scale dependence only set in at next-to-next-to-leading
order.
In order to assess the agreement between experimental data and theoretical
predictions quantitatively, we evaluate the $\chi^2$ values per degree of
freedom, $\chi_{\mathrm{DF}}^2$, for $\xi=1/2$, 1, and 2.
As in Ref.~\cite{kkp}, we select the data points with $x>0.1$.
The resulting $\chi^2_{\mathrm{DF}}$ values and the respective numbers of data
points are summarized in Table~\ref{tab:one}.
The $\chi_{\mathrm{DF}}^2$ values are of order unity, just as for the data
used in our fits \cite{kkp}.
We thus conclude that inclusion of the LEP2 data \cite{D2,O2} in our fits
would have had little impact on the resulting FFs.
For $\sqrt s=133$ and 161~GeV, the data are best described by the central
scale choice $\xi=1$, while the data at higher energies tend to prefer
$\xi=1/2$.
From the nice overall agreement we conclude that the predicted scaling
violations are confirmed by experiment.

\boldmath
\subsection{Hadroproduction in $p\bar p$ Collisions}
\unboldmath

The differential cross section
$Ed^3\sigma/d^3p=(1/2\pi p_T)d^2\sigma/dy\,dp_T$ of $p\bar p\to h^\pm+X$ was
measured at S$p\bar p$S and the Tevatron as a function of $p_T$ at various
values of $\sqrt s$ in rapidity ($y$) intervals constrained by the respective
detector geometry, mostly in the central region.
Specifically, UA1 provided data at $\sqrt s=200$ \cite{UA1b}, 500 \cite{UA1b},
540 \cite{UA1a}, 630 \cite{UA1c}, and 900~GeV \cite{UA1b}, UA2 at
$\sqrt s=540$~GeV \cite{UA2a,UA2b}, and CDF at $\sqrt s=630$ and 1800~GeV
\cite{CDF}.
This data is compared with our NLO predictions for $\xi=1/2$, 1, and 2 in 
Figs.~\ref{fig:u1a}--\ref{fig:c}.
The respective $\chi_{\mathrm{DF}}^2$ values are presented in
Table~\ref{tab:one}.
They are evaluated using the data points with $p_T>\mu_0/\xi$, where
$\mu_0=\sqrt2$~GeV is the starting scale of our FFs \cite{kkp}, so that
backward evolution is avoided.
For $\sqrt s$ values of order 500~GeV, the scale variation ranges from about
$\pm60\%$ at $p_T=3$~GeV to about $\pm25\%$ at $p_T=25$~GeV, and at a given 
value of $p_T$ it decreases with increasing values of $\sqrt s$.
The data is generally well described by our central predictions, both in
normalization and shape, even at low values of $p_T$, where corrections from
beyond NLO and nonperturbative effects are expected to be significant.
However, the UA1 data sets with $\sqrt s=200$ \cite{UA1b} and 630~GeV 
\cite{UA1c} and the UA2 data set with $1<|y|<1.8$ \cite{UA2b} tend to prefer
the scale choice $\xi=1/2$.
For the UA1 data at $\sqrt s=630$~GeV \cite{UA1c}, this was also noticed in
Ref.~\cite{bou}.
In that paper, it was also found that the overall agreement between this data
and the NLO prediction for $\xi=1/2$ does not improve if the normalization
factor for the gluon FF is modified within the tolerance allowed by the
original fit and the other FF parameters are in turn adjusted so as to
optimize the fit.
As for the UA2 data with $1<|y|<1.8$ \cite{UA2b} and the UA1 data at
$\sqrt s=630$~GeV \cite{UA1c}, we should also remark that the experimental
situation is somewhat ambiguous, since the UA1 data at $\sqrt s=540$~GeV 
\cite{UA1a} and the CDF data at $\sqrt s=630$~GeV clearly favour $\xi=1$.

At first sight, one might be worried by the relatively large
$\chi_{\mathrm{DF}}^2$ values obtained for $\xi=2$.
However, we have to bear in mind that, according to the prescription 
underlying Table~\ref{tab:one}, they are evaluated from all data points with
$p_T>\mu_0/2\approx0.71$~GeV.
On the one hand, the low-$p_T$ data from the hadron colliders come with very
small errors.
On the other hand, perturbation theory is known to break down in the low-$p_T$ 
limit.
Therefore, the $\chi_{\mathrm{DF}}^2$ values for $\xi=2$ in 
Table~\ref{tab:one} should be considered as exploratory.
In order to have a more conservative basis to judge the agreement between
experiment and theory and to facilitate the comparison between the cases
$\xi=1/2$, 1, and 2, we present in Table~\ref{tab:two} the corresponding
$\chi_{\mathrm{DF}}^2$ values obtained by selecting the data points with
$p_T>2\mu_0\approx2.83$~GeV, independent of $\xi$.
This happens at the expense of reducing the number of data points contributing
to the evaluation of the $\chi_{\mathrm{DF}}^2$ values.
In particular, the UA2 data with $|y|<0.63$ \cite{UA2a} is excluded
altogether, and just two out of the 22 CDF data points at $\sqrt s=630$~GeV
\cite{CDF} are left.
Comparing Table~\ref{tab:two} with Table~\ref{tab:one}, we observe that the
$\chi_{\mathrm{DF}}^2$ values for $\xi=1$ and 2 are considerably reduced,
especially for the UA1 data at $\sqrt s=630$~GeV \cite{UA1c} and the CDF data
at $\sqrt s=1800$~GeV \cite{CDF}.

\boldmath
\subsection{Photoproduction in $e^\pm p$ Collisions}
\unboldmath

H1 \cite{H1} (ZEUS \cite{ZEUS}) measured the differential cross section
$d^2\sigma/dy\,dp_T^2$ of inclusive charged-hadron photoproduction in
collisions of 27.5~GeV positrons (26.7~GeV electrons) and 820~GeV protons,
with $\sqrt s=300$~GeV (296~GeV), as a function of $p_T$ for $|y|<1$
($-1.2<y<1.4$) in the laboratory frame.
Here, $y$ is taken to be positive in the proton beam direction.
In this experiment, the scattered lepton was tagged, so that the virtuality
$Q^2$ of the Weizs\"acker-Williams photon was limited to be $Q^2<0.01$~GeV$^2$
(0.02~GeV$^2$), and the $\gamma p$ invariant mass $W$ was taken to be in the
interval $165<W<251$~GeV ($167<W<194$~GeV).
H1 also measured the $y$ distributions $d\sigma/dy$ for $p_T>2$ and 3~GeV.
In Figs.~\ref{fig:hp}--\ref{fig:z} this data is compared with our NLO
predictions for $\xi=1/2$, 1, and 2.
The scale variation approximately amounts to $\pm35\%$ at $p_T=3$~GeV and to
$\pm15\%$ at $p_T=12$~GeV.
The corresponding $\chi_{\mathrm{DF}}^2$ values are given in
Tables~\ref{tab:one} and \ref{tab:two}.
As in the case of the $p\bar p$ data, they are evaluated using the data points
with $p_T>\mu_0/\xi$ and $p_T>2\mu_0$, respectively.
In the case of the $y$ distribution for $p_T>2$~GeV, the NLO prediction for
$\xi=1/2$ also involves energy scales $\mu<\mu_0$ and should, therefore, be
taken with a grain of salt.
The corresponding $\chi_{\mathrm{DF}}^2$ value in Table~\ref{tab:one} is
marked by an asterisk.
We observe that the $p_T$ and $y$ distributions from H1 \cite{H1} are nicely
described by the theoretical predictions for $\xi=1$, as for both
normalization and shape, with $\chi_{\mathrm{DF}}^2\alt1$.
On the other hand, the ZEUS data \cite{ZEUS} prefer $\xi=2$ or even larger
values of $\xi$.

In Fig.~\ref{fig:hy}, we also investigate the dependence of the NLO prediction
on the photon PDFs.
We observe that the $y$ distributions evaluated with the GRV \cite{GRV} set
have a slightly larger normalization than those evaluated with the AFG
\cite{AFG} set, but they essentially have the same shape.
By contrast, the $y$ distributions based on the GS \cite{GS} set are
appreciably steeper.
However, we should keep in mind that the these differences are smaller than
the theoretical uncertainty due to scale variations.

\boldmath
\subsection{Photoproduction in $e^+e^-$ Collisions}
\unboldmath

OPAL \cite{Ogg} measured the differential cross section of
$e^+e^-\to e^+e^-h^\pm+X$ via two-photon collisions at LEP2, with average CM
energy $\sqrt s=166.5$~GeV, in the $\gamma\gamma$-invariant-mass interval
$10<W<125$~GeV and in the three subintervals $10<W<30$~GeV, $30<W<55$~GeV, and
$55<W<125$~GeV.
The scattered electrons and positrons were antitagged, the maximum scattering 
angles being $\theta^\prime=33$~mrad.
Specifically, they presented the $p_T$ distributions $d\sigma/dp_T$ for
$|y|<1.5$ and the $|y|$ distributions $d\sigma/d|y|$ for $p_T>1.5$~GeV.
Notice that the $y$ distributions are symmetric w.r.t.\ $y=0$, so that
$d\sigma/d|y|=2d\sigma/dy$.
This data is compared with our NLO predictions for $\xi=1/2$, 1, and 2 in 
Figs.~\ref{fig:op1} and \ref{fig:oy1}, respectively.
In the case of the calculation for $10<W<125$~GeV, the scale variation
approximately amounts to $\pm20\%$ at $p_T=3$~GeV and to $\pm5\%$ at
$p_T=15$~GeV.
The corresponding $\chi_{\mathrm{DF}}^2$ values are given in
Tables~\ref{tab:one} and \ref{tab:two}.
As in the case of the $p\bar p$ and $\gamma p$ data, they are evaluated using
the data points with $p_T>\mu_0/\xi$ and $p_T>2\mu_0$, respectively.
In Fig.~\ref{fig:oy1}, the NLO predictions for $\xi=1/2$ involve energy scales
$\mu<\mu_0$, as is indicated in Table~\ref{tab:one} by asterisks, but they are
nevertheless shown for illustration.
From the steep fall-off of the $p_T$ distributions in Fig.~\ref{fig:op1} it
follows that the $y$ distributions in Fig.~\ref{fig:oy1} are greatly dominated
by the low-$p_T$ range, where the theoretical uncertainty due to scale
variations is considerably larger than the experimental errors and 
nonperturbative effects may be significant.
Thus, the $p_T$ cut-off in Fig.~\ref{fig:oy1} is likely to be too small for
perturbation theory to be meaningful.
We observe that the $p_T$ and $y$ distributions are nicely described by the
NLO predictions for $\xi=1$, as for both normalization and shape.

In Figs.~\ref{fig:op2} and \ref{fig:oy2}, we illustrate how our default NLO
predictions (evaluated with the AFG \cite{AFG} photon PDFs for $\xi=1$) for
$10<W<125$~GeV shown in Figs.~\ref{fig:op1} and \ref{fig:oy1}, respectively,
are built up by their direct, single-resolved, and double-resolved components.
As expected, the double-resolved channel dominates at low values of $p_T$,
while the direct channel wins out at large values of $p_T$.
However, one should keep in mind that this decomposition depends on the scale
choice and the factorization scheme.
In Figs.~\ref{fig:op2} and \ref{fig:oy2}, we also analyze the theoretical 
uncertainty related to the photon PDFs.
Similarly to Fig.~\ref{fig:hy}, the NLO predictions based on the AFG
\cite{AFG} and GRV \cite{GRV} sets only differ in their overall
normalizations.
By contrast, the $p_T$ and $y$ distributions evaluated with the GS \cite{GS}
set exhibit a somewhat different shape.
There is actually a minor caveat attached to the latter: the interval of $p_T$
integration involves values below the starting scale $\mu_0=\sqrt3$~GeV of the
GS \cite{GS} set.
The comments concerning the validity of perturbation theory made in connection
with Fig.~\ref{fig:oy1} also apply to Fig.~\ref{fig:oy2}.
We conclude that, except in the low-$p_T$ range, the theoretical uncertainty
due to the photon PDFs is insignificant, of order 10\% for $p_T>4$~GeV.

\section{Comparisons with Other FF Sets
\label{sec:three}}

In order to compare the three NLO FF sets that were released earlier this year 
\cite{kkp,kre,bou}, we consider the differential cross section
$(1/\sigma_{\mathrm{tot}})d\sigma/dx$ of inclusive charged-hadron production
from the fragmentation of light ($uds$), $c$, $b$, and all quarks in $e^+e^-$
annihilation at $\sqrt s=29$, 91.2, and 189~GeV.
We compare the resulting NLO predictions for $\xi=1$ with PEP \cite{T1},
LEP1 \cite{D,D1}, SLC \cite{S}, and LEP2 \cite{O2} data.
In contrast to Ref.~\cite{bou}, we refrain from scale optimization.
Specifically, we consider the all-flavour samples from DELPHI \cite{D1}, SLD 
\cite{S}, and OPAL \cite{O2}, the light-quark samples from TPC \cite{T1} and
DELPHI \cite{D,D1}, the $c$-quark samples from TPC \cite{T1} and DELPHI 
\cite{D}, and the $b$-quark samples from TPC \cite{T1} and DELPHI \cite{D,D1}.
Notice that the DELPHI analysis of Ref.~\cite{D} is based on data collected
during the years 1991--1993, while the one of Ref.~\cite{D1} is based on 1994
data.
The outcome of these comparisons is presented in
Figs.~\ref{fig:c1}--\ref{fig:c3}, which refer to $\sqrt s=29$, 91.2, and 
189~GeV, respectively, and Table~\ref{tab:three}.

In order to properly assess these comparisons, it is useful to recall which of
these data samples actually entered the fits of Refs.~\cite{kkp,kre,bou}.
The flavour-specific data samples from TPC \cite{T1} were used in
Ref.~\cite{kre}, the 1991--1993 data from DELPHI \cite{D} in Ref.~\cite{bou},
the 1994 data from DELPHI \cite{D1} in Refs.~\cite{kkp,bou},\footnote{%
The light-quark sample from Ref.~\cite{D1} was not used in Ref.~\cite{bou}.}
and the SLD \cite{S} data in Refs.~\cite{kkp,kre,bou}.
For each of the three analyses \cite{kkp,kre,bou}, the data samples which were 
not used as input are marked in Table~\ref{tab:three} by asterisks.
We did not include the flavour-specific data samples from TPC \cite{T1} in our
fit \cite{kkp} because they are unpublished.
Furthermore, we only incorporated the most recent data from DELPHI \cite{D1}.
In contrast to Refs.~\cite{bkk,kre,bou}, we did not make use of the ALEPH
\cite{Al,A} data, for reasons explained in Ref.~\cite{kkp}.\footnote{%
Due to a misinterpretation of Refs.~\cite{Al,A}, the respective
$\chi_{\mathrm{DF}}^2$ values for $\sigma^h(\mathrm{all})$, $\sigma^h(uds)$,
$\sigma^h(c)$, and $\sigma^h(b)$ quoted in Table~1 of Ref.~\cite{kkp} are
incorrect.
The correct numbers read 48.1, 11.1, 7.61, and 104, respectively.}

From Figs.~\ref{fig:c1}--\ref{fig:c3}, we observe that the three NLO FF sets
\cite{kkp,kre,bou} yield very similar all-flavour and light-quark results, 
except for the small-$x$ region ($x\alt0.1$), where the QCD-improved parton
model is known to be unreliable, and the large-$x$ region ($x\agt0.8$), where
the experimental data is less constraining anyway.
On the other hand, there are appreciable differences in the $c$- and $b$-quark 
contributions also at intermediate values of $x$.
In the $c$-quark case, the K \cite{kre} result undershoots the KKP \cite{kkp}
and BFGW \cite{bou} ones, while the KKP \cite{kkp} result undershoots the K
\cite{kre} and BFGW \cite{bou} ones in the $b$-quark case.
The latter feature may be understood by observing that the $b$-quark FFs of 
Ref.~\cite{kkp} were determined so as to fit the more recent DELPHI \cite{D1}
data, whereas, for example, the BFGW \cite{bou} analysis used the older DELPHI
\cite{D} data as input.
As may be seen from Fig.~\ref{fig:c2}, the $b$-quark data samples of these two 
measurements \cite{D,D1} are not compatible with each other.
This was also observed in Ref.~\cite{bou}.
Thus, it is impossible to simultaneously describe both $b$-quark data samples 
\cite{D,D1} with $\chi_{\mathrm{DF}}^2$ values of order unity, as is also
evident from Table~\ref{tab:three}.
Leaving aside this $b$-quark anomaly, we conclude from Table~\ref{tab:three}
that the three NLO FF sets \cite{kkp,kre,bou} are compatible with each other 
and nicely describe the considered data samples \cite{D,D1,S,T1}, with 
$\chi_{\mathrm{DF}}^2$ values of order unity, even in those cases where the 
latter were not included in the fits.

In this section, we compared the cross sections of charged-hadron production
from the fragmentation of light, $c$, $b$, and all quarks resulting from the
three NLO FF sets \cite{kkp,kre,bou} rather than the individual FFs at some
particular value of $\mu^2$.
Comparisons of the latter kind were performed in Refs.~\cite{kre,bou} taking 
into account the old BKK \cite{bkk} sets rather than the new KKP \cite{kkp} 
ones, and appreciable differences were found, in particular at large values of 
$x$.
We feel that, for the time being, such comparisons are of limited usefulness,
especially as far as the individual $u$-, $d$-, and $s$-quark FFs are 
concerned.
These FFs are not separately constrained by available $e^+e^-$ data, and
different theoretical assumptions were made in Refs.~\cite{kkp,kre,bou}.

\section{Conclusions
\label{sec:four}}

\begin{figure}[ht]
\begin{center}
\centerline{\epsfig{figure=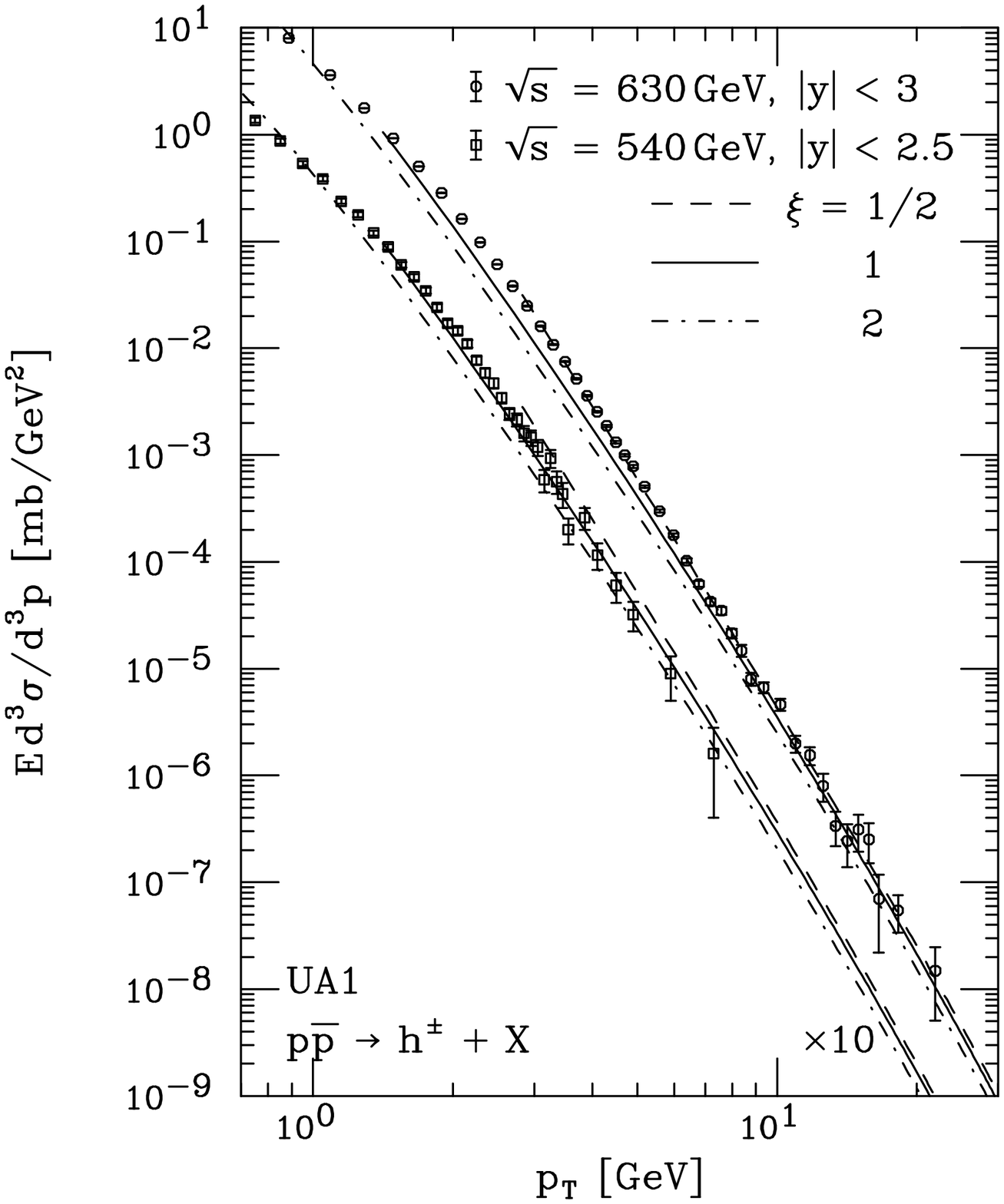,height=16cm}}
\caption{Differential cross section $Ed^3\sigma/d^3p$ (in mb/GeV$^2$) of
inclusive charged-hadron hadroproduction in $p\bar p$ collisions
$p\bar p\to h^\pm+X$ as a function of transverse momentum $p_T$ at CM energies
$\protect\sqrt s=540$ and 630~GeV, averaged over rapidity intervals $|y|<2.5$
and 3, respectively.
The NLO predictions for $\xi=1/2$ (dashed lines), 1 (solid lines), and 2 
(dot-dashed lines) are compared with data from UA1 \protect\cite{UA1a,UA1c}.
The lower set of curves is rescaled by a factor of 1/10.}
\label{fig:u1a}
\end{center}
\end{figure}

\begin{figure}[ht]
\begin{center}
\centerline{\epsfig{figure=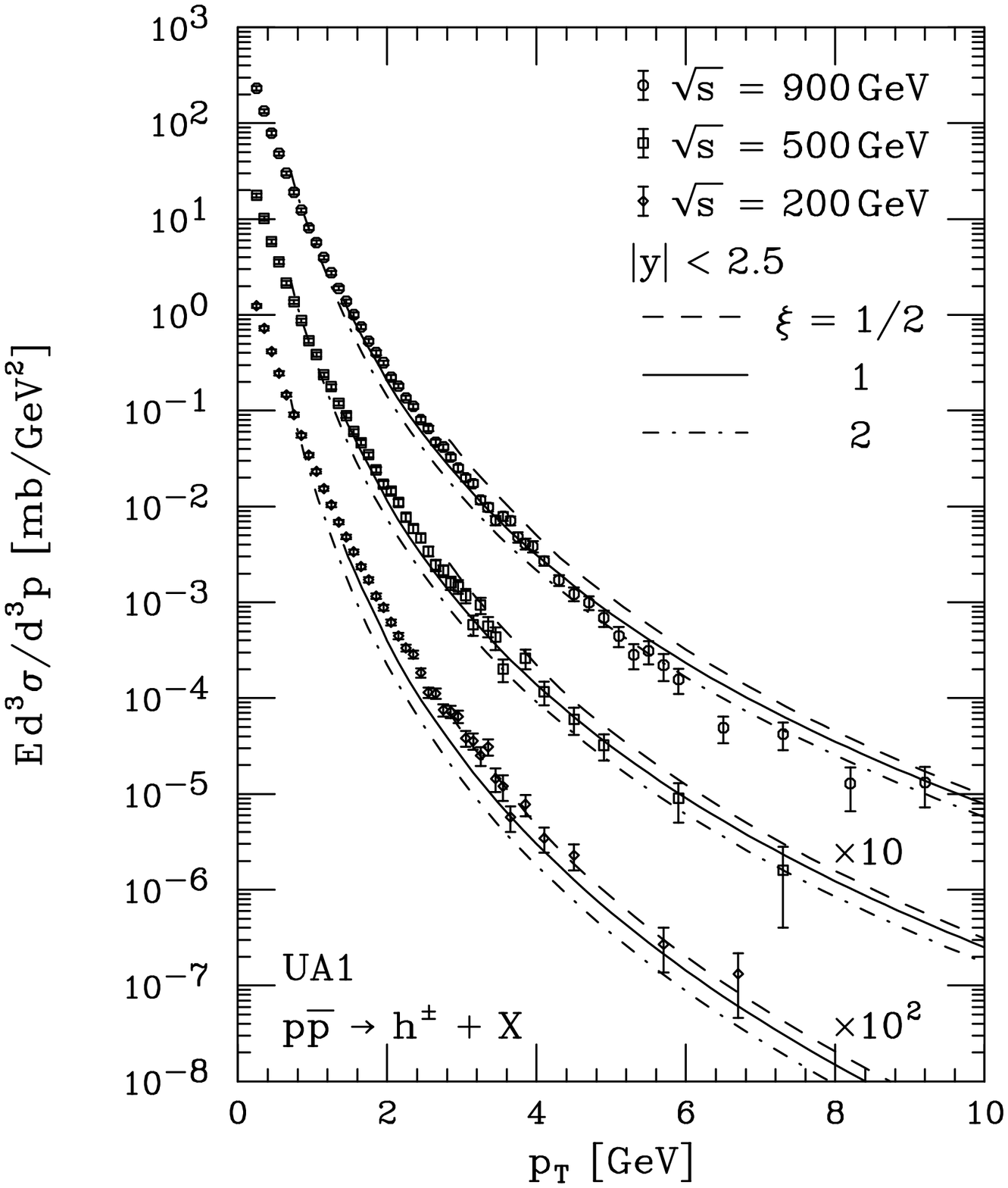,height=16cm}}
\caption{Differential cross section
$Ed^3\sigma/d^3p$ (in mb/GeV$^2$) of inclusive charged-hadron hadroproduction
in $p\bar p$ collisions $p\bar p\to h^\pm+X$ as a function of transverse
momentum $p_T$ at CM energies $\protect\sqrt s=200$, 500, and 900~GeV,
averaged over rapidity interval $|y|<2.5$.
The NLO predictions for $\xi=1/2$ (dashed lines), 1 (solid lines), and 2 
(dot-dashed lines) are compared with data from UA1 \protect\cite{UA1b}.
Each set of curves is rescaled relative to the nearest upper one by a factor
of 1/10.}
\label{fig:u1b}
\end{center}
\end{figure}

\begin{figure}[ht]
\begin{center}
\centerline{\epsfig{figure=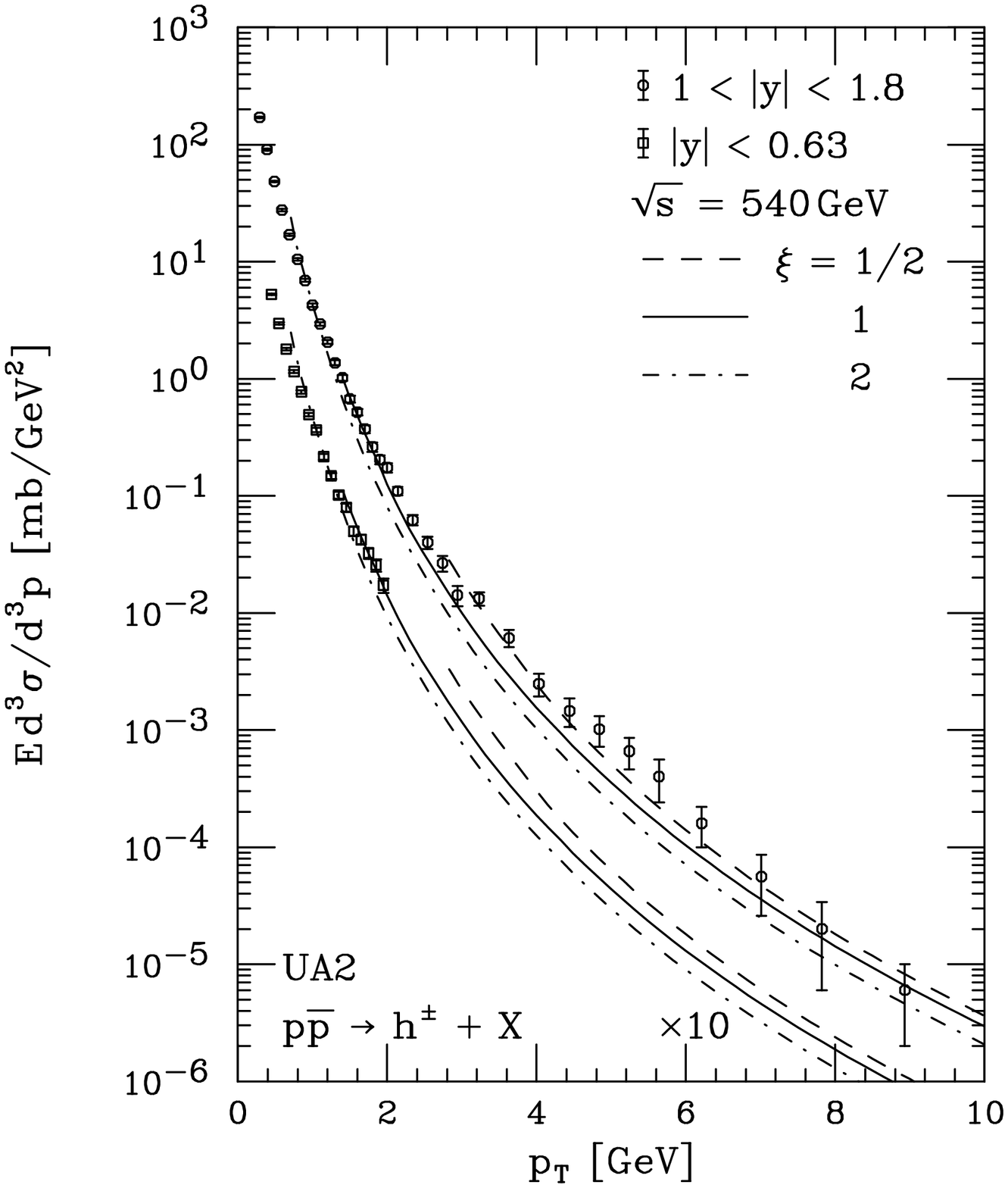,height=16cm}}
\caption{Differential cross section
$Ed^3\sigma/d^3p$ (in mb/GeV$^2$) of inclusive charged-hadron hadroproduction
in $p\bar p$ collisions $p\bar p\to h^\pm+X$ as a function of transverse
momentum $p_T$ at CM energy $\protect\sqrt s=540$~GeV, averaged over
rapidity intervals $|y|<0.63$ and $1<|y|<1.8$.
The NLO predictions for $\xi=1/2$ (dashed lines), 1 (solid lines), and 2 
(dot-dashed lines) are compared with data from UA2 \protect\cite{UA2a,UA2b}.
The lower set of curves is rescaled by a factor of 1/10.}
\label{fig:u2}
\end{center}
\end{figure}

\begin{figure}[ht]
\begin{center}
\centerline{\epsfig{figure=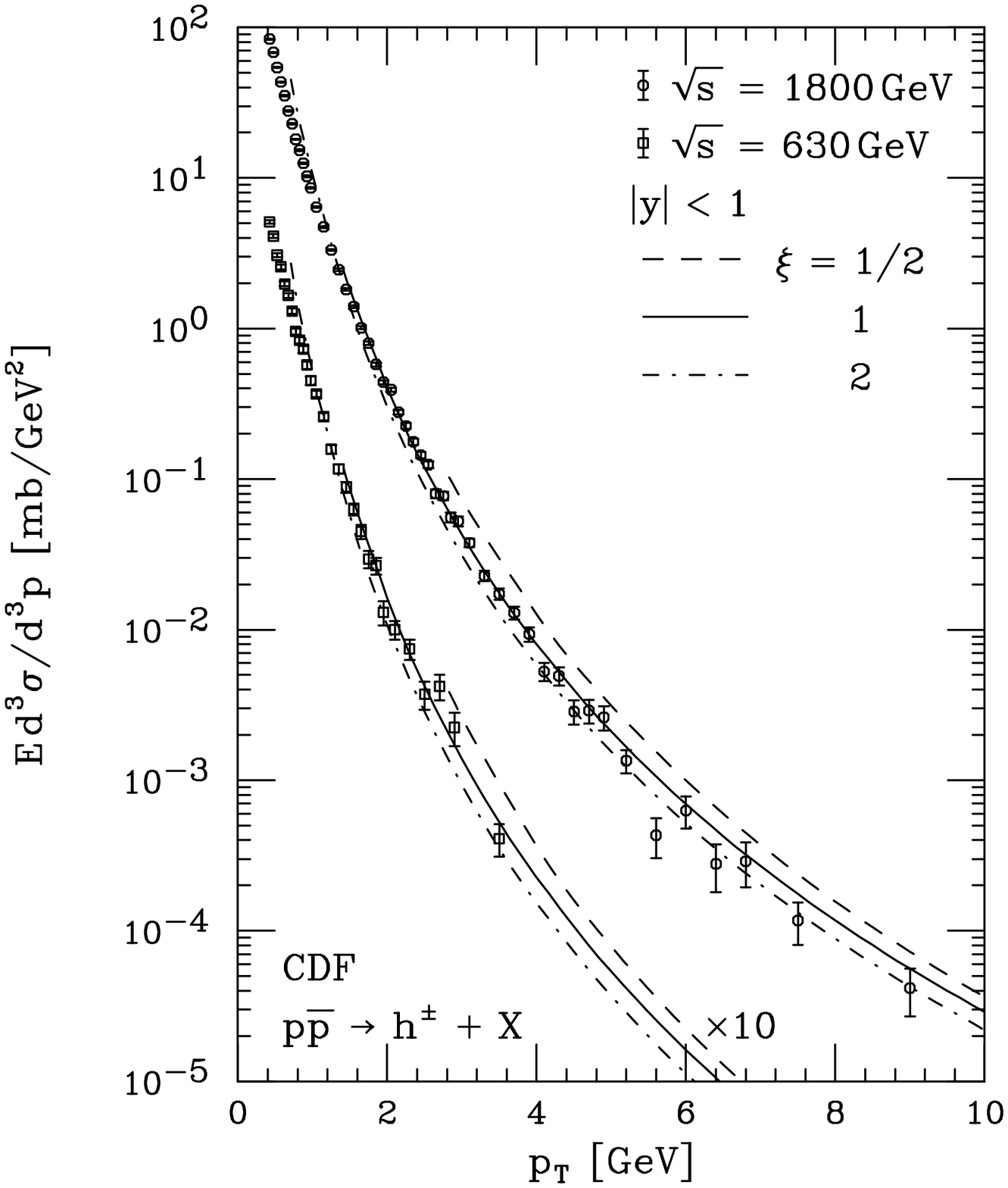,height=16cm}}
\caption{Differential cross section $Ed^3\sigma/d^3p$ (in mb/GeV$^2$) of
inclusive charged-hadron hadroproduction in $p\bar p$ collisions
$p\bar p\to h^\pm+X$ as a function of transverse momentum $p_T$ at CM energies
$\protect\sqrt s=630$ and 1800~GeV, averaged over rapidity interval $|y|<1$.
The NLO predictions for $\xi=1/2$ (dashed lines), 1 (solid lines), and 2 
(dot-dashed lines) are compared with data from CDF \protect\cite{CDF}.
The lower set of curves is rescaled by a factor of 1/10.}
\label{fig:c}
\end{center}
\end{figure}

\begin{figure}[ht]
\begin{center}
\centerline{\epsfig{figure=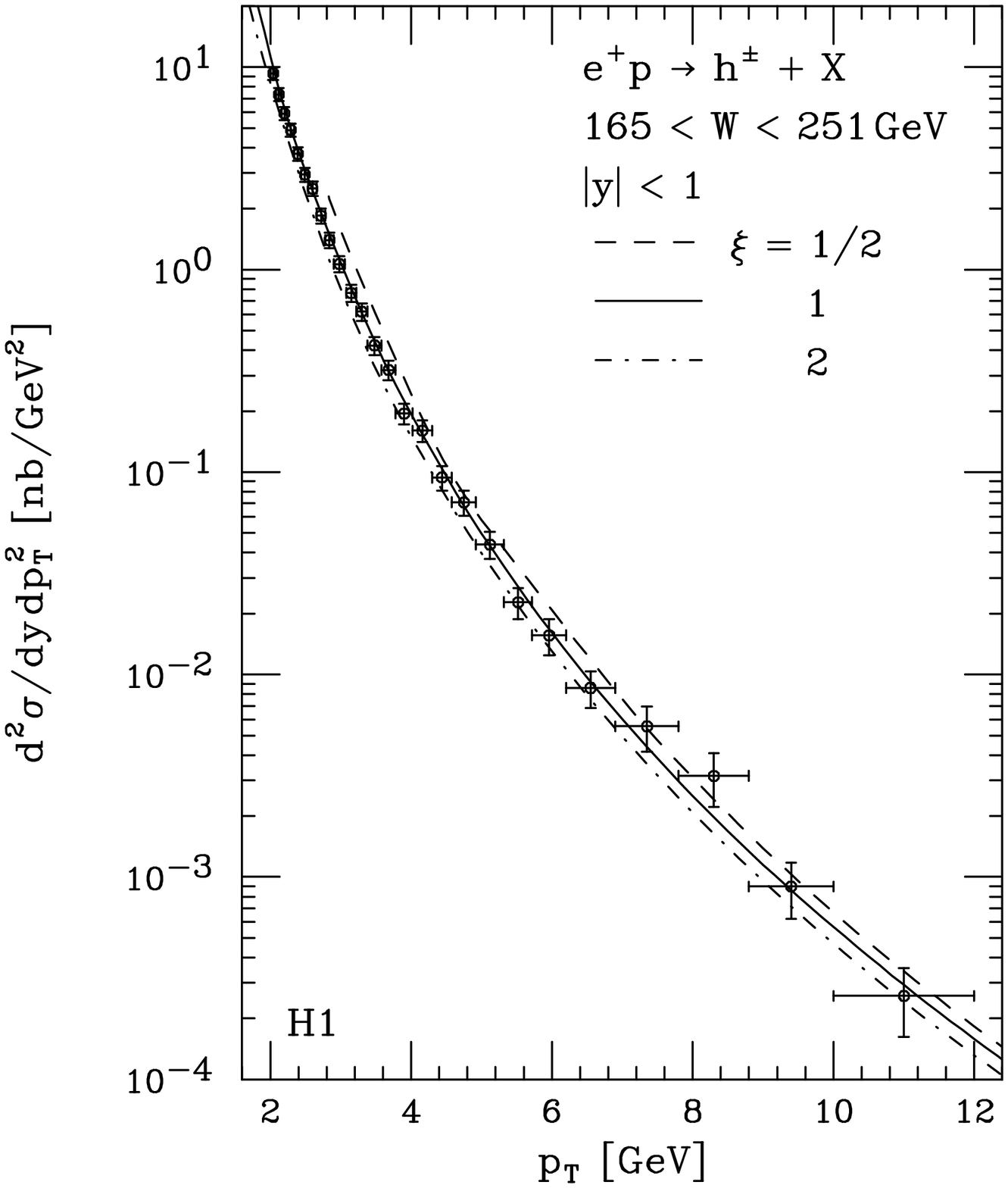,height=16cm}}
\caption{Differential cross section $d^2\sigma/dy\,dp_T^2$ (in nb/GeV$^2$) of
inclusive charged-hadron photoproduction in $e^+p$ collisions
$e^+p\to h^\pm+X$ as a function of transverse momentum $p_T$ at CM energy
$\protect\sqrt s=300$~GeV, integrated over $\gamma p$-invariant-mass interval
$165<W<251$~GeV and averaged over rapidity interval $|y|<1$ in the laboratory
frame.
The scattered positron is tagged, the maximum photon virtuality being
$Q_{\mathrm{max}}^2=0.01$~GeV$^2$.
The NLO predictions evaluated with the AFG \protect\cite{AFG} photon PDFs for
$\xi=1/2$ (dashed lines), 1 (solid lines), and 2 (dot-dashed lines) are
compared with data from H1 \protect\cite{H1}.}
\label{fig:hp}
\end{center}
\end{figure}

\begin{figure}[ht]
\begin{center}
\centerline{\epsfig{figure=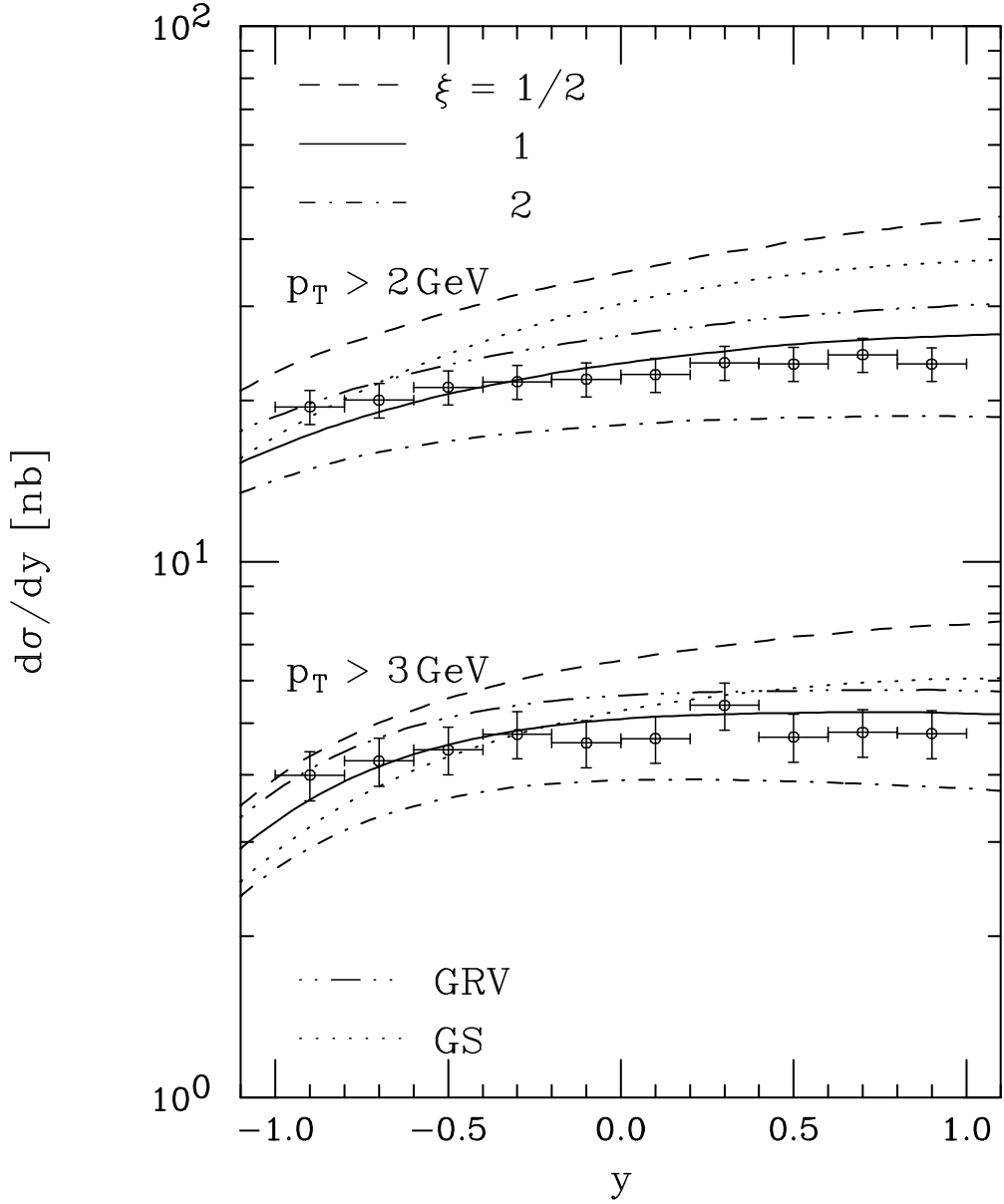,height=16cm}}
\caption{Differential cross section $d\sigma/dy$ (in nb) of inclusive
charged-hadron photoproduction in $e^+p$ collisions $e^+p\to h^\pm+X$ as a
function of rapidity $y$ at CM energy $\protect\sqrt s=300$~GeV, integrated
over $\gamma p$-invariant-mass interval $165<W<251$~GeV and
transverse-momentum intervals $2<p_T<12$~GeV and $3<p_T<12$~GeV.
The scattered positron is tagged, the maximum photon virtuality being
$Q_{\mathrm{max}}^2=0.01$~GeV$^2$.
The NLO predictions evaluated with the AFG \protect\cite{AFG} photon PDFs
for $\xi=1/2$ (dashed lines), 1 (solid lines), and 2 (dot-dashed lines) are
compared with data from H1 \protect\cite{H1}.
The dot-dot-dashes and dotted lines correspond to the GRV \protect\cite{GRV}
and GS \protect\cite{GS} photon PDFs.}
\label{fig:hy}
\end{center}
\end{figure}

\begin{figure}[ht]
\begin{center}
\centerline{\epsfig{figure=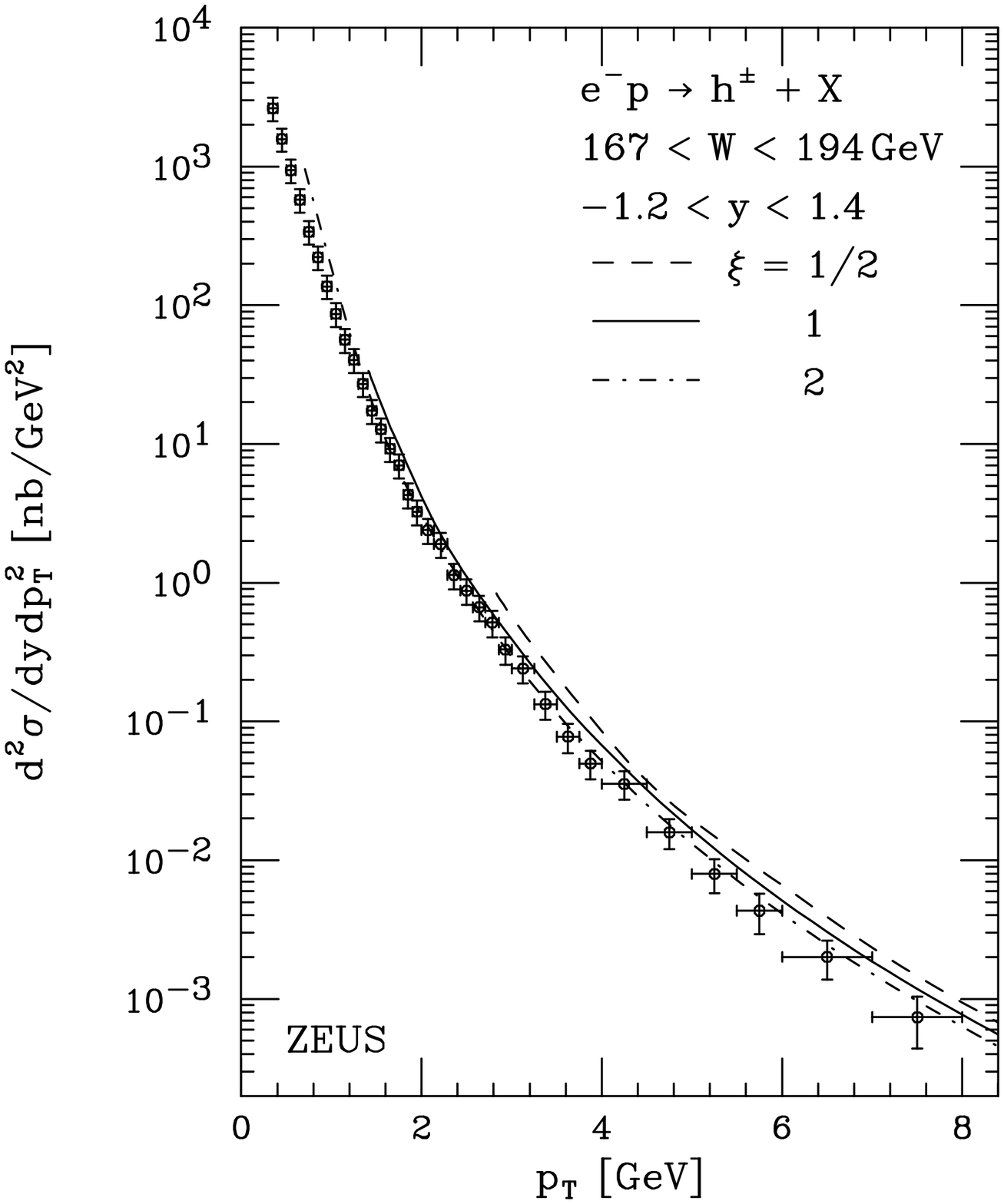,height=16cm}}
\caption{Differential cross section $d^2\sigma/dy\,dp_T^2$ (in nb/GeV$^2$) of
inclusive charged-hadron photoproduction in $e^-p$ collisions
$e^-p\to h^\pm+X$ as a function of transverse momentum $p_T$ at CM energy
$\protect\sqrt s=296$~GeV, integrated over $\gamma p$-invariant-mass interval
$167<W<194$~GeV and averaged over rapidity interval
$-1.2<y<1.4$ in the laboratory frame.
The scattered electron is tagged, the maximum photon virtuality being
$Q_{\mathrm{max}}^2=0.02$~GeV$^2$.
The NLO predictions evaluated with the AFG \protect\cite{AFG} photon PDFs for
$\xi=1/2$ (dashed lines), 1 (solid lines), and 2 (dot-dashed lines) are
compared with data from ZEUS \protect\cite{ZEUS}.}
\label{fig:z}
\end{center}
\end{figure}

\begin{figure}[ht]
\begin{center}
\centerline{\epsfig{figure=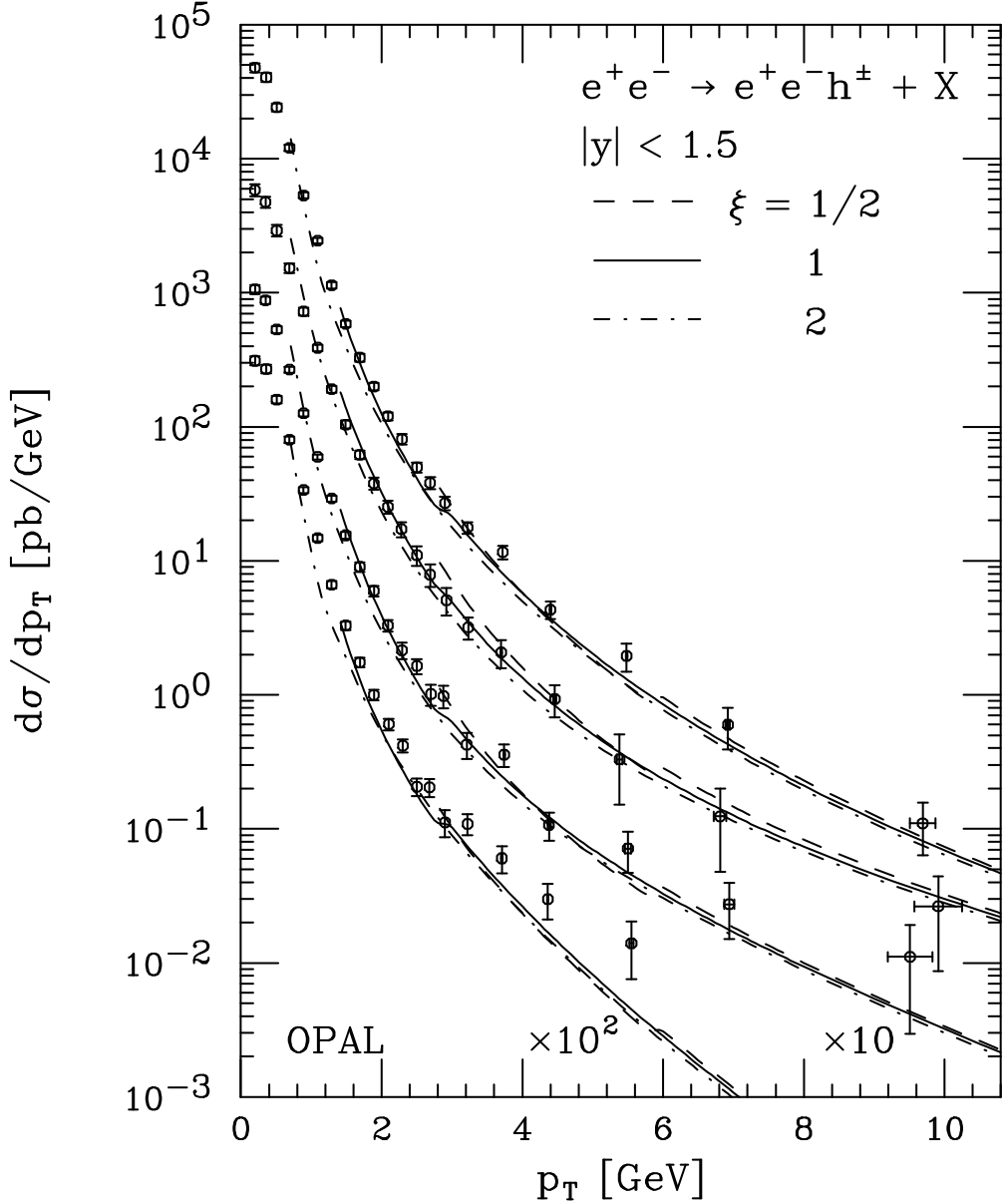,height=16cm}}
\caption{Differential cross section $d\sigma/dp_T$ (in pb/GeV) of inclusive
charged-hadron photoproduction in $\gamma\gamma$ collisions
$e^+e^-\to e^+e^-h^\pm+X$ as a function of transverse momentum $p_T$ at CM
energy $\protect\sqrt s=166.5$~GeV, integrated over
$\gamma\gamma$-invariant-mass intervals $10<W<30$~GeV, $30<W<55$~GeV,
$55<W<125$~GeV, and $10<W<125$~GeV (from bottom to top in this order) and
rapidity interval $|y|<1.5$.
The scattered electrons and positrons are antitagged, the maximum scattering 
angles being $\theta^\prime=33$~mrad.
The NLO predictions evaluated with the AFG \protect\cite{AFG} photon PDFs for
$\xi=1/2$ (dashed lines), 1 (solid lines), and 2 (dot-dashed lines) are
compared with data from OPAL \protect\cite{Ogg}.
The lowest two sets of curves are rescaled relative to the nearest upper one
by a factor of 1/10.}
\label{fig:op1}
\end{center}
\end{figure}

\begin{figure}[ht]
\begin{center}
\centerline{\epsfig{figure=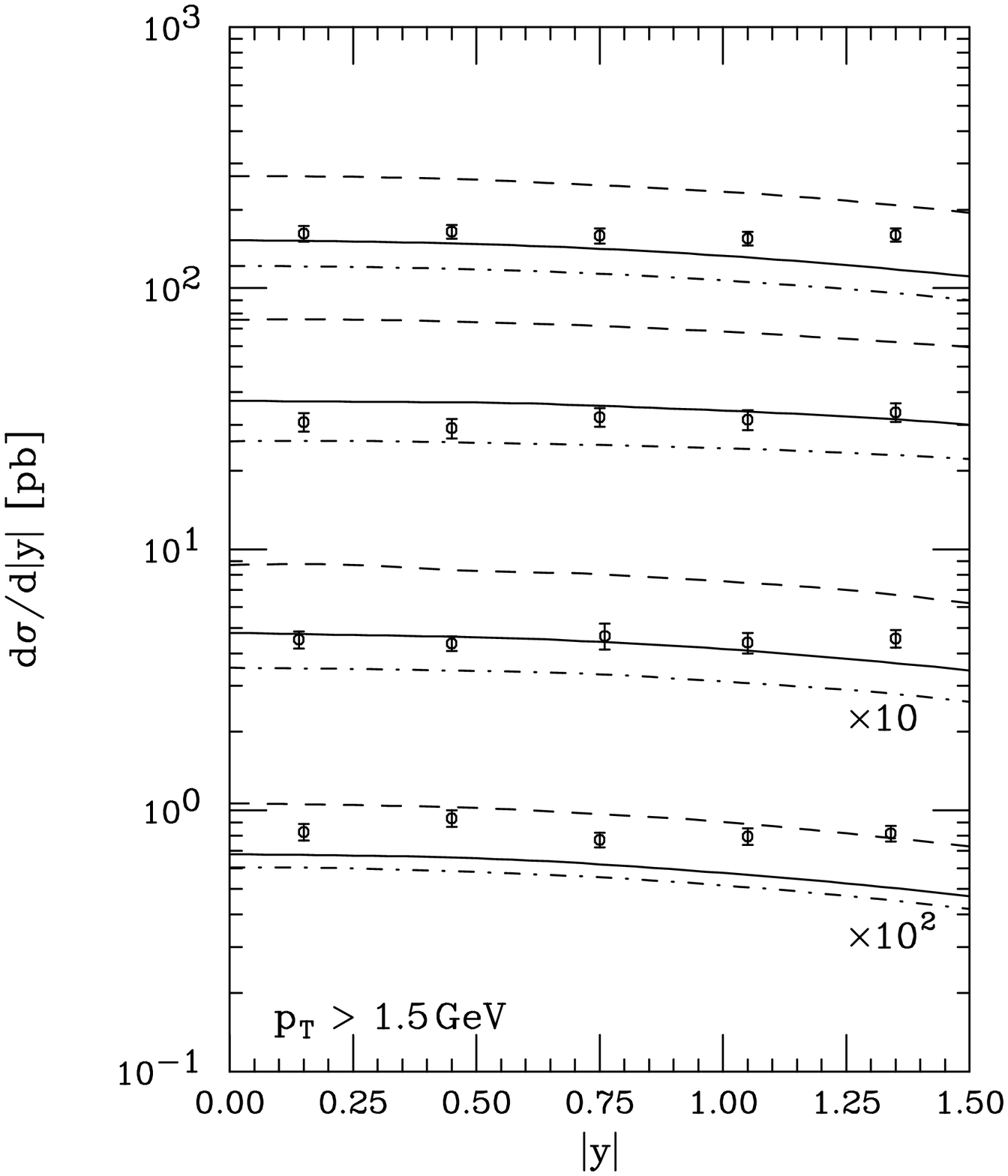,height=16cm}}
\caption{Differential cross section $d\sigma/d|y|$ (in pb) of inclusive
charged-hadron photoproduction in $\gamma\gamma$ collisions
$e^+e^-\to e^+e^-h^\pm+X$ as a function of rapidity $|y|$ at CM energy
$\protect\sqrt s=166.5$~GeV, integrated over $\gamma\gamma$-invariant-mass
intervals $10<W<30$~GeV, $30<W<55$~GeV, $55<W<125$~GeV, and $10<W<125$~GeV
(from bottom to top in this order) and transverse-momentum interval
$1.5<p_T<15$~GeV.
The scattered electrons and positrons are antitagged, the maximum scattering 
angles being $\theta^\prime=33$~mrad.
The NLO predictions evaluated with the AFG \protect\cite{AFG} photon PDFs for
$\xi=1/2$ (dashed lines), 1 (solid lines), and 2 (dot-dashed lines) are
compared with data from OPAL \protect\cite{Ogg}.
The lowest two sets of curves are rescaled relative to the nearest upper one
by a factor of 1/10.}
\label{fig:oy1}
\end{center}
\end{figure}

\begin{figure}[ht]
\begin{center}
\centerline{\epsfig{figure=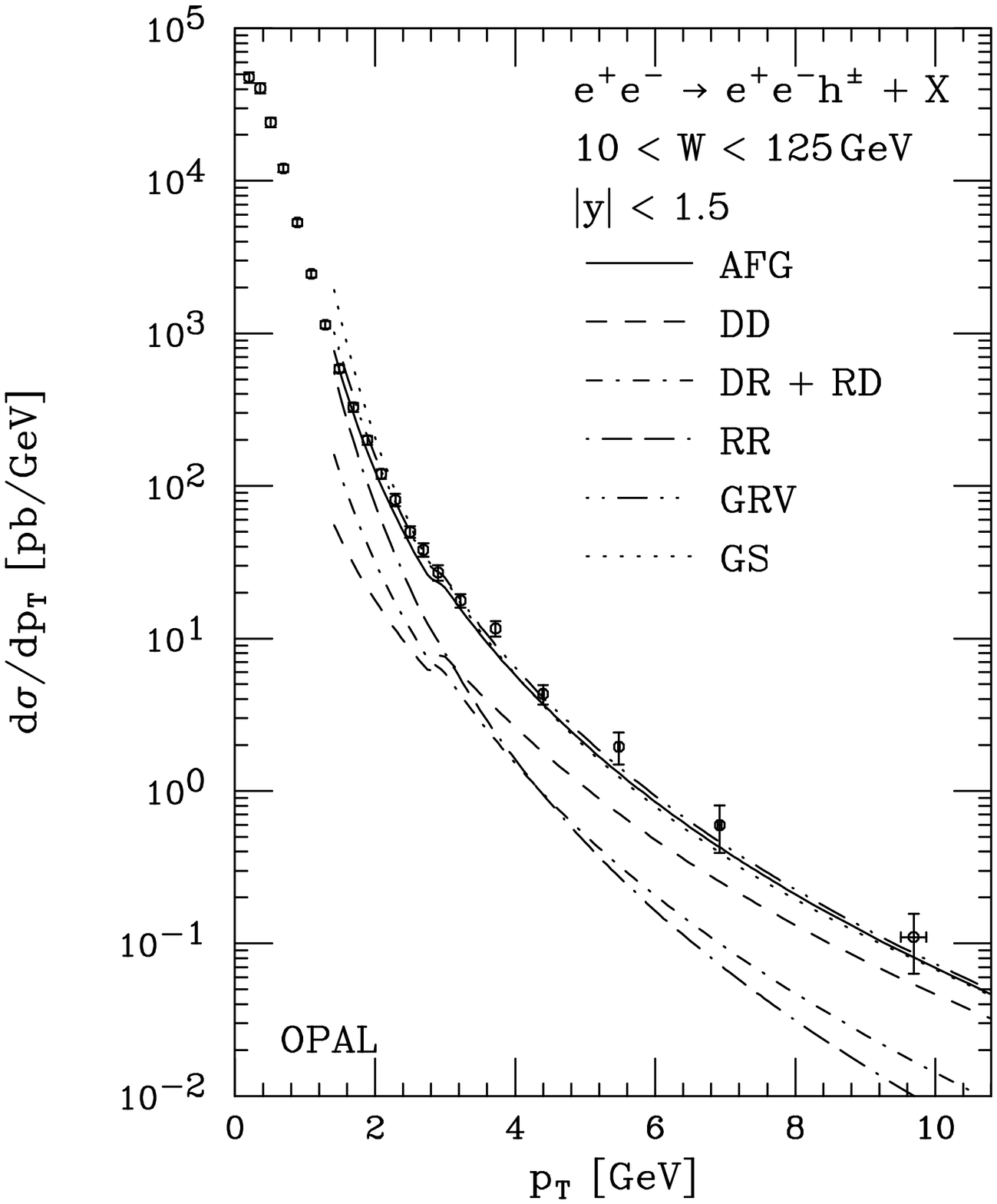,height=16cm}}
\caption{Differential cross section $d\sigma/dp_T$ (in pb/GeV) of inclusive
charged-hadron photoproduction in $\gamma\gamma$ collisions
$e^+e^-\to e^+e^-h^\pm+X$ as a function of transverse momentum $p_T$ at CM
energy $\protect\sqrt s=166.5$~GeV, integrated over 
$\gamma\gamma$-invariant-mass interval $10<W<125$~GeV and rapidity interval
$|y|<1.5$.
The scattered electrons and positrons are antitagged, the maximum scattering 
angles being $\theta^\prime=33$~mrad.
The NLO predictions evaluated with the AFG \protect\cite{AFG} (solid line),
GRV \protect\cite{GRV} (dot-dot-dashed line), and GS \protect\cite{GS} (dotted
line) photon PDFs for $\xi=1$ are compared with data from OPAL
\protect\cite{Ogg}.
For comparison, also the direct (dashed line), single-resolved (dot-dashed 
line), and double-resolved (dash-dash-dotted line) components of the AFG
prediction are shown.}
\label{fig:op2}
\end{center}
\end{figure}

\begin{figure}[ht]
\begin{center}
\centerline{\epsfig{figure=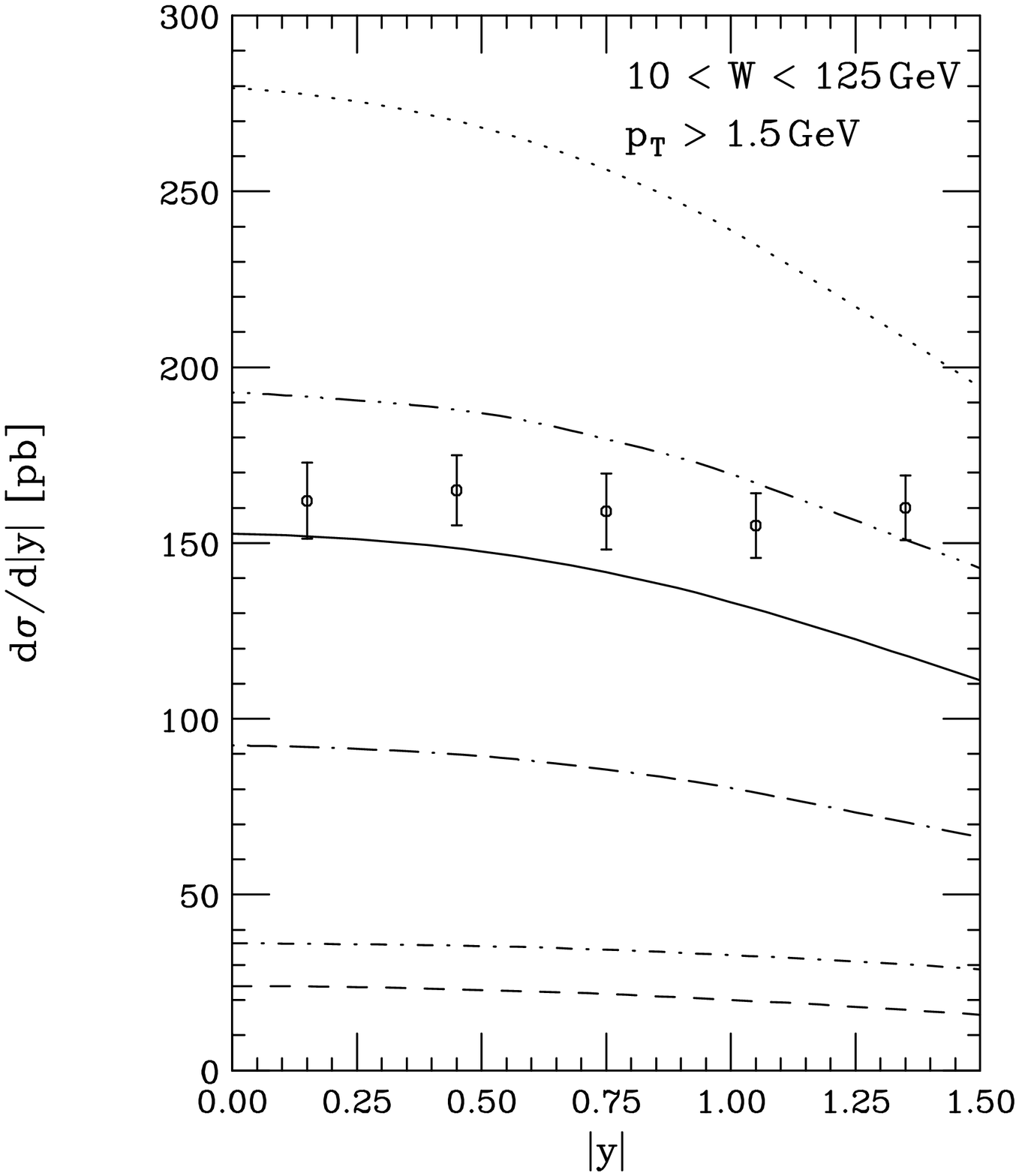,height=16cm}}
\caption{Differential cross section $d\sigma/d|y|$ (in pb) of inclusive
charged-hadron photoproduction in $\gamma\gamma$ collisions
$e^+e^-\to e^+e^-h^\pm+X$ as a function of rapidity $|y|$ at CM
energy $\protect\sqrt s=166.5$~GeV, integrated over 
$\gamma\gamma$-invariant-mass interval $10<W<125$~GeV and transverse-momentum
interval $1.5<p_T<15$~GeV.
The scattered electrons and positrons are antitagged, the maximum scattering 
angles being $\theta^\prime=33$~mrad.
The NLO predictions evaluated with the AFG \protect\cite{AFG} (solid line),
GRV \protect\cite{GRV} (dot-dot-dashed line), and GS \protect\cite{GS} (dotted
line) photon PDFs for $\xi=1$ are compared with data from OPAL
\protect\cite{Ogg}.
For comparison, also the direct (dashed line), single-resolved (dot-dashed 
line), and double-resolved (dash-dash-dotted line) components of the AFG
prediction are shown.}
\label{fig:oy2}
\end{center}
\end{figure}

\begin{figure}[ht]
\begin{center}
\centerline{\epsfig{figure=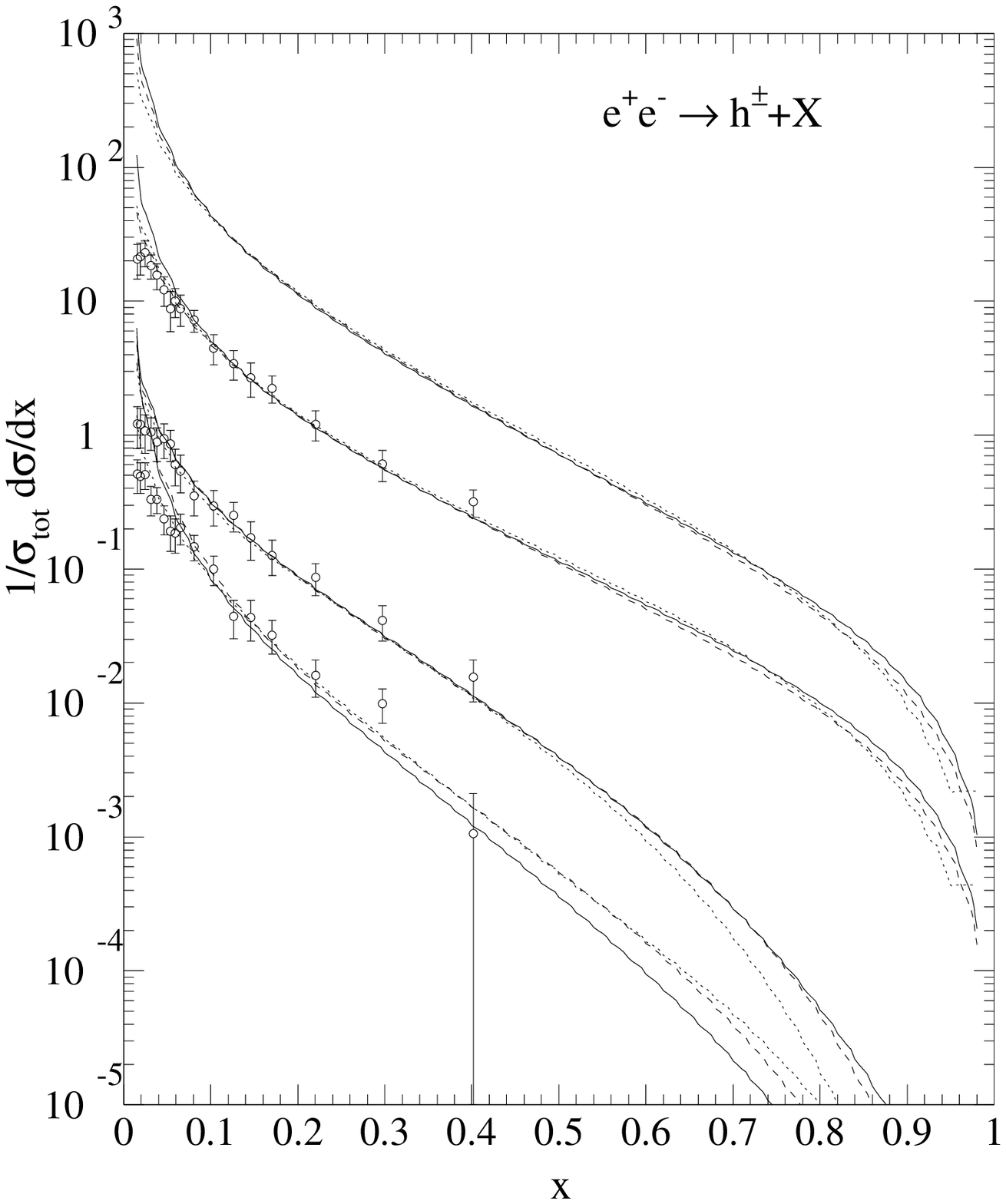,height=16cm}}
\caption{Normalized differential cross section
$(1/\sigma_{\mathrm{tot}})d\sigma/dx$ (in pb) of inclusive charged-hadron
production from $b$, $c$, light, and all quarks (from bottom to top in this
order) in $e^+e^-$ annihilation $e^+e^-\to h^\pm+X$ as a function of scaled
momentum $x$ at CM energy $\protect\sqrt s=29$~GeV.
The NLO predictions for $\xi=1$ based on the KKP \protect\cite{kkp} (solid
lines), K \protect\cite{kre} (dotted lines), and BFGW \protect\cite{bou}
(dashed lines) FFs are compared with data from TPC \protect\cite{T1}.
Each set of curves is rescaled relative to the nearest upper one by a factor
of 1/5.}
\label{fig:c1}
\end{center}
\end{figure}

\begin{figure}[ht]
\begin{center}
\centerline{\epsfig{figure=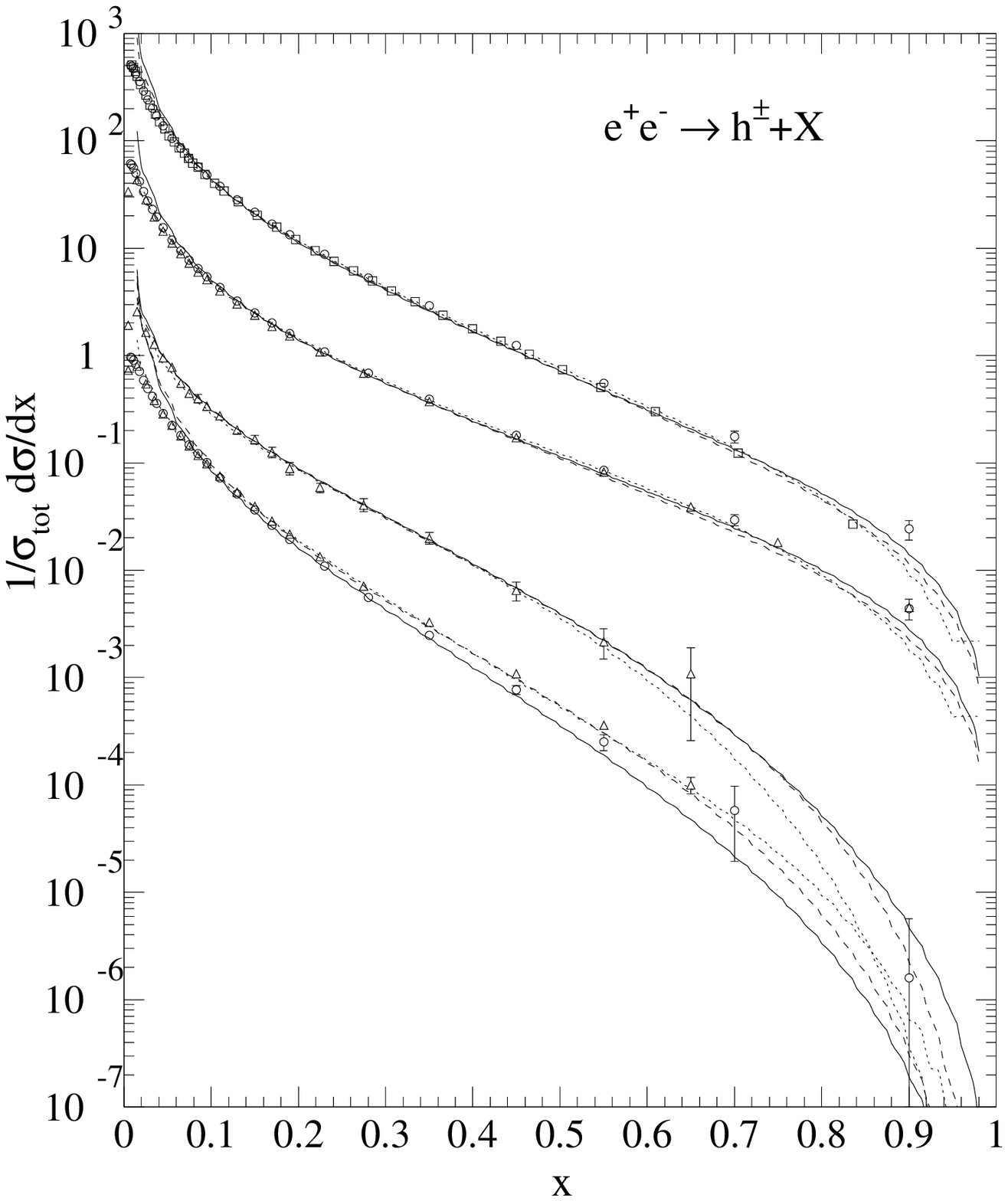,height=16cm}}
\caption{Normalized differential cross section
$(1/\sigma_{\mathrm{tot}})d\sigma/dx$ (in pb) of inclusive charged-hadron
production from $b$, $c$, light, and all quarks (from bottom to top in this
order) in $e^+e^-$ annihilation $e^+e^-\to h^\pm+X$ as a function of scaled
momentum $x$ at CM energy $\protect\sqrt s=91.2$~GeV.
The NLO predictions for $\xi=1$ based on the KKP \protect\cite{kkp} (solid
lines), K \protect\cite{kre} (dotted lines), and BFGW \protect\cite{bou}
(dashed lines) FFs are compared with data from DELPHI (yields of 1991--1993
(triangles) \protect\cite{D} and 1994 (circles) \protect\cite{D1}) and SLD
(squares) \protect\cite{S}.
Each set of curves is rescaled relative to the nearest upper one by a factor
of 1/5.}
\label{fig:c2}
\end{center}
\end{figure}

\begin{figure}[ht]
\begin{center}
\centerline{\epsfig{figure=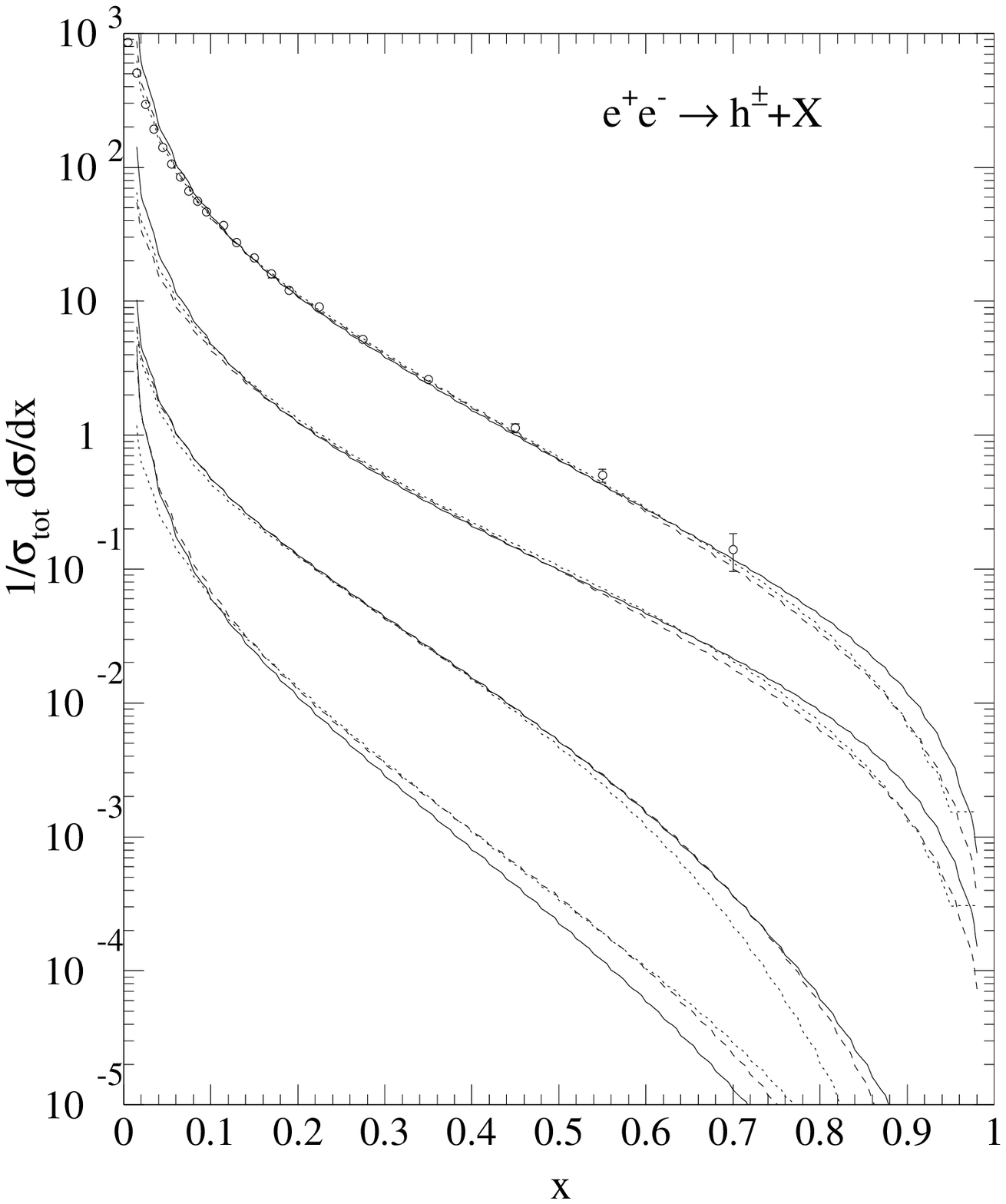,height=16cm}}
\caption{Normalized differential cross section
$(1/\sigma_{\mathrm{tot}})d\sigma/dx$ (in pb) of inclusive charged-hadron
production from $b$, $c$, light, and all quarks (from bottom to top in this
order) in $e^+e^-$ annihilation $e^+e^-\to h^\pm+X$ as a function of scaled
momentum $x$ at CM energy $\protect\sqrt s=189$~GeV.
The NLO predictions for $\xi=1$ based on the KKP \protect\cite{kkp} (solid
lines), K \protect\cite{kre} (dotted lines), and BFGW \protect\cite{bou}
(dashed lines) FFs are compared with data from OPAL \protect\cite{O2}.
Each set of curves is rescaled relative to the nearest upper one by a factor
of 1/5.}
\label{fig:c3}
\end{center}
\end{figure}

Although the FFs $D_a^h(x,\mu^2)$ characterizing the hadronization of partons
$a$ into light hadrons $h$ are genuinely nonperturbative objects, they possess
two important properties that follow from perturbative considerations within
the QCD-improved parton model and are amenable to experimental tests.
Firstly, their dependence on the factorization scale $\mu$ is supposed to be
determined by the timelike Altarelli-Parisi equations \cite{gri}.
Secondly, the factorization theorem predicts that they are universal in the
sense that they only depend on the fragmenting parton $a$ and the inclusively
produced hadron $h$, but not on the details of the process from which they
were determined.
In this paper, we tested these two properties at NLO for the $\pi^\pm$,
$K^\pm$, and $p/\bar p$ FFs recently fitted \cite{kkp} to experimental data of
$e^+e^-$ annihilation from LEP1 \cite{A1,D1,Ag}, SLC \cite{S}, and PEP
\cite{T}.
On the one hand, we confronted new data on inclusive charged-hadron production
from LEP2 \cite{D2,O2}, with CM energies ranging from 133~GeV to 189~GeV, with
our NLO predictions.
On the other hand, we performed a global NLO analysis of essentially all 
high-statistics data on inclusive charged-hadron production in colliding-beam
experiments, including $p\bar p$ scattering at $Sp\bar pS$
\cite{UA1a,UA1b,UA1c,UA2a,UA2b} and the Tevatron \cite{CDF}, $\gamma p$
scattering at HERA \cite{H1,ZEUS}, and $\gamma\gamma$ scattering at LEP2
\cite{Ogg}.
In all cases, we found reasonable agreement between the experimental data and 
our NLO predictions as for both normalization and shape, as may be seen from
Figs.~\ref{fig:do}--\ref{fig:oy2} and Tables~\ref{tab:one} and \ref{tab:two}.
The majority of the data sets are best described with the central scale choice
$\xi=1$.
Exceptions include the UA1 data sets with $\sqrt s=200$ \cite{UA1b} and
630~GeV \cite{UA1c} and the UA2 data set with $1<|y|<1.8$ \cite{UA2b}, which
prefer $\xi=1/2$, as well as the ZEUS data, which favours $\xi=2$.
However, if we estimate the theoretical uncertainty due to unknown corrections
beyond NLO by varying $\xi$ between 1/2 and 2, as is usually done, then it is
justified to state that all the considered data sets agree with our NLO
predictions within their errors.
We hence conclude that our global analysis of inclusive charged-hadron 
production provides evidence that both the predicted scaling violations and
the universality of the FFs are realized in nature.

As is well known, the gluon FF enters the prediction for the unpolarized cross
section $d\sigma/dx$ of inclusive hadron production in $e^+e^-$ annihilation
only at NLO, while at LO it only contributes indirectly via the $\mu^2$
evolution.
In order to nevertheless have a handle on it, we included in our fits
\cite{kkp} experimental data on gluon-tagged three-jet events from LEP1
\cite{Ag}.
Furthermore, we checked that our predictions for the longitudinal cross
section, where it already enters at LO, agree well with available data
\cite{Al,Dl}.
On the other hand, the gluon FF is known to play a crucial r\^ole for
$p\bar p$, $\gamma p$, and $\gamma\gamma$ scattering at low values of $p_T$.
Thus, the comparisons performed here provide another nontrivial test of the
gluon FF.

As we have seen in Figs.~\ref{fig:u1a}--\ref{fig:hp}, \ref{fig:z}, and 
\ref{fig:op1}, the theoretical uncertainty of the NLO predictions due to 
scale variations significantly decreases as $p_T$ increases.
In order to perform more meaningful comparisons, it would, therefore, be
desirable if $p\bar p$, $\gamma p$, and $\gamma\gamma$ experiments extended
their measurements out to larger values of $p_T$.
Furthermore, in order to render such comparisons more specific, it would be
useful if these experiments provided us with separate data samples of
$\pi^\pm$, $K^\pm$, and $p/\bar p$ hadrons.

We also estimated the current systematic uncertainties in the FFs by comparing
our NLO predictions for flavour-tagged inclusive charged-hadron production
with those obtained from two other FF sets which appeared during this year
\cite{kre,bou}.
Apart from a difference in the $b$-quark FFs at medium to large values of $x$,
which may be traced to the incompatibility of two underlying
$b$-quark-specific data samples \cite{D,D1}, all three FF sets 
\cite{kkp,kre,bou} mutually agree within the present experimental errors.
We conclude that the determination of NLO FFs from global fits to experimental 
data has reached a level of precision which is comparable to the one familiar
from similar analyses for PDFs.

\vspace{1cm}
\noindent
{\bf Acknowledgements}
\smallskip

\noindent
We thank S. Kretzer for providing us with the data of Ref.~\cite{T1} in 
numerical form.
The II. Institut f\"ur Theoretische Physik is supported in part by the
Deutsche Forschungsgemeinschaft through Grant No.\ KN~365/1-1, by the
Bundesministerium f\"ur Bildung und Forschung through Grant No.\ 05~HT9GUA~3,
and by the European Commission through the Research Training Network
{\it Quantum Chromodynamics and the Deep Structure of Elementary Particles}
under Contract No.\ ERBFMRX-CT98-0194.

\begin{appendix}

\section{Parameterizations}

For the reader's convenience, we present here simple parameterizations of the
$x$ and $\mu^2$ dependences of our LO and NLO sets of $\pi^\pm$, $K^\pm$, and
$p/\bar p$ FFs \cite{kkp}.
As in Ref.~\cite{bkk}, we introduce the scaling variable
\begin{equation}
\bar s=\ln\frac{\ln\left(\mu^2/\Lambda_{\overline{\mathrm{MS}}}^{(5)}\right)}
{\ln\left(\mu_0^2/\Lambda_{\overline{\mathrm{MS}}}^{(5)}\right)}.
\end{equation}
In LO (NLO), we have $\Lambda_{\overline{\mathrm{MS}}}^{(5)}=88$~MeV (213~MeV)
\cite{kkp}.
We use three different values for $\mu_0$, namely \cite{kkp}
\begin{eqnarray}
\mu_0&=&\left\{\begin{array}{l@{\quad\mbox{if}\quad}l}
\sqrt2~\mbox{GeV}, & a=u,d,s,g\\
m(\eta_c)=2.9788~\mbox{GeV}, & a=c\\
m(\Upsilon)=9.46037~\mbox{GeV}, & a=b
\end{array}\right..
\end{eqnarray}
This leads to three different definitions of $\bar s$.
For definiteness, we use the symbol $\bar s_c$ for charm and $\bar s_b$ for
bottom along with $\bar s$ for the residual partons.
We parameterize our FFs as
\begin{equation}
\label{temp}
D(x,\mu^2)=Nx^\alpha(1-x)^\beta\left(1+\frac{\gamma}{x}\right)
\end{equation}
and express the coefficients $N$, $\alpha$, $\beta$, and $\gamma$ as
polynomials in $\bar s$, $\bar s_c$, and $\bar s_b$.
For $\bar s=\bar s_c=\bar s_b=0$, the parameterizations agree with Eq.~(2) of
Ref.~\cite{kkp} in combination with the appropriate entries in Table~2 of that
paper.
The charm and bottom parameterizations must be put to zero by hand for
$\bar s_c<0$ and $\bar s_b<0$, respectively.

We list below the parameters to be inserted in Eq.~(\ref{temp}).
The resulting parameterizations describe the properly evolved FFs to an 
accuracy of 10\% or less for $\mu_0<\mu<200$~GeV and $0.1<x<0.8$.
Deviations in excess of 10\% may occur outside these regions.
\begin{description}
\item LO FFs for $(\pi^++\pi^-)$:
\begin{description}
\item $D_u^{\pi^\pm}(x,\mu^2)=D_d^{\pi^\pm}(x,\mu^2)$:
\begin{eqnarray}
N&=&0.54610-0.22946\bar s-0.22594\bar s^2+0.21119\bar s^3\nonumber\\
\alpha&=&-1.46616-0.45404\bar s-0.12684\bar s^2+0.27646\bar s^3\nonumber\\
\beta&=&1.01864+0.95367\bar s-1.09835\bar s^2+0.74657\bar s^3\nonumber\\
\gamma&=&-0.01877\bar s+0.02949\bar s^2
\end{eqnarray}
\item $D_s^{\pi^\pm}(x,\mu^2)$:
\begin{eqnarray}
N&=&22.2815-20.8125\bar s-11.5725\bar s^2+15.5372\bar s^3\nonumber\\
\alpha&=&0.12732+0.23075\bar s-2.71424\bar s^2+1.72456\bar s^3\nonumber\\
\beta&=&6.13697+2.18849\bar s-5.04475\bar s^2+3.29117\bar s^3\nonumber\\
\gamma&=&0.09044\bar s-0.07589\bar s^2 
\end{eqnarray}
\item $D_c^{\pi^\pm}(x,\mu^2)$:
\begin{eqnarray}
N&=&8.75500-9.32277\bar s_c+1.80600\bar s_c^2+2.02179\bar s_c^3\nonumber\\
\alpha&=&-0.38611-0.41190\bar s_c-0.48496\bar s_c^2+0.42525\bar s_c^3
\nonumber\\
\beta&=&5.61846+0.74035\bar s_c-0.64929\bar s_c^2+0.66788\bar s_c^3\nonumber\\
\gamma&=&0.06652\bar s_c-0.05531\bar s_c^2
\end{eqnarray}
\item $D_b^{\pi^\pm}(x,\mu^2)$:
\begin{eqnarray}
N&=&0.31147-0.19319\bar s_b-0.10487\bar s_b^2+0.18824\bar s_b^3\nonumber\\
\alpha&=&-1.92993-0.44692\bar s_b-0.08271\bar s_b^2+0.30441\bar s_b^3 
\nonumber\\
\beta&=&3.47086+0.79775\bar s_b-0.28091\bar s_b^2+0.39504\bar s_b^3\nonumber\\
\gamma&=&-0.04887\bar s_b+0.03212\bar s_b^2
\end{eqnarray}
\item $D_g^{\pi^\pm}(x,\mu^2)$:
\begin{eqnarray}
N&=&6.04510-6.61523\bar s-1.64978\bar s^2+2.68223\bar s^3\nonumber\\
\alpha&=&-0.71378+0.14705\bar s-1.08423\bar s^2-0.43182\bar s^3\nonumber\\
\beta&=&2.92133+1.48429\bar s+1.32887\bar s^2-1.78696\bar s^3\nonumber\\
\gamma&=&0.23086\bar s-0.29182\bar s^2
\end{eqnarray}
\end{description}
\item NLO FFs for $(\pi^++\pi^-)$:
\begin{description}
\item $D_u^{\pi^\pm}(x,\mu^2)=D_d^{\pi^\pm}(x,\mu^2)$:
\begin{eqnarray}
N&=&0.44809-0.13828\bar s-0.06951\bar s^2+0.01354\bar s^3\nonumber\\
\alpha&=&-1.47598-0.30498\bar s-0.01863\bar s^2-0.12529\bar s^3\nonumber\\
\beta&=&0.91338+0.64145\bar s+0.07270\bar s^2-0.16989\bar s^3\nonumber\\
\gamma&=&0.07396\bar s-0.07757\bar s^2
\end{eqnarray}
\item $D_s^{\pi^\pm}(x,\mu^2)$:
\begin{eqnarray}
N&=&16.5987-18.3856\bar s+2.44225\bar s^2+2.13225\bar s^3\nonumber\\
\alpha&=&0.13345+0.22712\bar s-0.83625\bar s^2+0.38526\bar s^3\nonumber\\
\beta&=&5.89903-0.16911\bar s+0.59886\bar s^2-0.25630\bar s^3\nonumber\\
\gamma&=& -0.18619\bar s+0.87362\bar s^2 
\end{eqnarray}
\item $D_c^{\pi^\pm}(x,\mu^2)$:
\begin{eqnarray}
N&=&6.17173-4.82450\bar s_c-1.30844\bar s_c^2+1.95527\bar s_c^3\nonumber\\
\alpha&=&-0.53618-0.27879\bar s_c-0.51337\bar s_c^2+0.10900\bar s_c^3
\nonumber\\
\beta&=&5.60108+0.83571\bar s_c-1.15141\bar s_c^2+0.77027\bar s_c^3\nonumber\\
\gamma&=&0.09268\bar s_c-0.11267\bar s_c^2
\end{eqnarray}
\item $D_b^{\pi^\pm}(x,\mu^2)$:
\begin{eqnarray}
N&=&0.25944-0.11499\bar s_b+0.03733\bar s_b^2-0.18028\bar s_b^3\nonumber\\
\alpha&=&-1.98713-0.35858\bar s_b+0.22277\bar s_b^2-0.66413\bar s_b^3
\nonumber\\
\beta&=&3.52857+0.72303\bar s_b+0.46260\bar s_b^2 -0.99235\bar s_b^3
\nonumber\\
\gamma&=&-0.02701\bar s_b-0.02089\bar s_b^2
\end{eqnarray}
\item $D_g^{\pi^\pm}(x,\mu^2)$:
\begin{eqnarray}
N&=&3.73331-3.16946\bar s-0.47683\bar s^2+0.70270\bar s^3\nonumber\\
\alpha&=&-0.74159-0.51377\bar s-0.19705\bar s^2-0.17917\bar s^3\nonumber\\
\beta&=&2.33092+2.03394\bar s-0.50764\bar s^2-0.08565\bar s^3\nonumber\\
\gamma&=& 0.09466\bar s-0.10222\bar s^2
\end{eqnarray}
\end{description}
\item LO FFs for $(K^++K^-)$:
\begin{description}
\item $D_u^{K^\pm}(x,\mu^2)=D_s^{K^\pm}(x,\mu^2)$:
\begin{eqnarray}
N&=&0.25937-0.10502\bar s+0.00572\bar s^2-0.00269\bar s^3\nonumber\\
\alpha&=&-0.61925+0.09956\bar s+0.07389\bar s^2-0.00070\bar s^3\nonumber\\
\beta&=&0.85946+0.57965\bar s+0.26397\bar s^2-0.12764\bar s^3\nonumber\\
\gamma&=&0.15303\bar s+0.14807\bar s^2
\end{eqnarray}
\item $D_d^{K^\pm}(x,\mu^2)$:
\begin{eqnarray}
N&=&5.38115-3.05084\bar s-1.10056\bar s^2+1.31207\bar s^3\nonumber\\
\alpha&=&-0.00321-0.25889\bar s-0.18494\bar s^2+0.13994\bar s^3\nonumber\\
\beta&=&3.07632+1.13745\bar s-0.90413\bar s^2+0.56581\bar s^3\nonumber\\
\gamma&=&0.05141\bar s-0.00697\bar s^2
\end{eqnarray}
\item $D_c^{K^\pm}(x,\mu^2)$:
\begin{eqnarray}
N&=&5.18266-3.48519\bar s_c-1.00982\bar s_c^2+1.17996\bar s_c^3\nonumber\\
\alpha&=&-0.17751+0.02309\bar s_c-0.61327\bar s_c^2-0.03532\bar s_c^3
\nonumber\\
\beta&=&4.30306+1.00547\bar s_c-0.51779\bar s_c^2+0.20683\bar s_c^3\nonumber\\
\gamma&=&0.13514\bar s_c-0.17778\bar s_c^2
\end{eqnarray}
\item $D_b^{K^\pm}(x,\mu^2)$:
\begin{eqnarray}
N&=&1.57044-1.78340\bar s_b+0.57100\bar s_b^2+0.15469\bar s_b^3\nonumber\\
\alpha&=&-0.84143-0.43448\bar s_b-0.05314\bar s_b^2-0.36621\bar s_b^3
\nonumber\\
\beta&=&6.01488+0.72953\bar s_b-0.64433\bar s_b^2+0.92351\bar s_b^3\nonumber\\
\gamma&=&0.01024\bar s_b-0.06160\bar s_b^2
\end{eqnarray}
\item $D_g^{K^\pm}(x,\mu^2)$:
\begin{eqnarray}
N&=&0.02862-0.02113\bar s+0.00389\bar s^2+0.00901\bar s^3\nonumber\\
\alpha&=&-2.94091+0.66881\bar s-0.29670\bar s^2+0.20574\bar s^3\nonumber\\
\beta&=&2.73474-0.58222\bar s+0.04329\bar s^2+0.78033\bar s^3\nonumber\\
\gamma&=& 0.03586\bar s-0.01220\bar s^2
\end{eqnarray}
\end{description}
\item NLO FFs for $(K^++K^-)$:
\begin{description}
\item $D_u^{K^\pm}(x,\mu^2)=D_s^{K^\pm}(x,\mu^2)$:
\begin{eqnarray}
N&=&0.17806-0.10988\bar s-0.02524\bar s^2+0.03142\bar s^3\nonumber\\
\alpha&=&-0.53733-0.60058\bar s+0.07863\bar s^2+0.13276\bar s^3\nonumber\\
\beta&=&0.75940+0.61356\bar s-0.43886\bar s^2+0.23942\bar s^3\nonumber\\
\gamma&=&0.10742\bar s+0.12800\bar s^2 
\end{eqnarray}
\item $D_d^{K^\pm}(x,\mu^2)$:
\begin{eqnarray}
N&=&4.96269+1.54098\bar s-9.06376\bar s^2+4.94791\bar s^3\nonumber\\
\alpha&=&0.05562+1.88660\bar s-2.94350\bar s^2+1.04227\bar s^3\nonumber\\
\beta&=&2.79926+3.02991\bar s-4.14807\bar s^2+1.91494\bar s^3\nonumber\\
\gamma&=&0.85450\bar s-0.61016\bar s^2 
\end{eqnarray}
\item $D_c^{K^\pm}(x,\mu^2)$:
\begin{eqnarray}
N&=&4.25954-5.44309\bar s_c+6.11031\bar s_c^2-3.13973\bar s_c^3\nonumber\\
\alpha&=&-0.24144-1.07757\bar s_c+1.52364\bar s_c^2-0.74308\bar s_c^3
\nonumber\\
\beta&=&4.21265+0.25590\bar s_c+0.98423\bar s_c^2-0.52893\bar s_c^3\nonumber\\
\gamma&=&-0.04000\bar s_c+0.08695\bar s_c^2
\end{eqnarray}
\item $D_b^{K^\pm}(x,\mu^2)$:
\begin{eqnarray}
N&=&1.32443-1.41156\bar s_b-0.04809\bar s_b^2+0.79066\bar s_b^3\nonumber\\
\alpha&=&-0.88351-0.44818\bar s_b-0.60073\bar s_b^2+0.45526\bar s_b^3 
\nonumber\\
\beta&=&6.15221+0.46679\bar s_b-0.50792\bar s_b^2+0.67006\bar s_b^3\nonumber\\
\gamma&=&-0.00477\bar s_b-0.05503\bar s_b^2
\end{eqnarray}
\item $D_g^{K^\pm}(x,\mu^2)$:
\begin{eqnarray}
N&=&0.23140-0.33644\bar s+0.16204\bar s^2-0.02598\bar s^3\nonumber\\
\alpha&=&-1.36400+0.97182\bar s-0.02908\bar s^2-0.43195\bar s^3\nonumber\\
\beta&=&1.79761+1.57116\bar s+0.71847\bar s^2-0.68331\bar s^3\nonumber\\
\gamma&=&0.36906\bar s+2.39060\bar s^2
\end{eqnarray}
\end{description}
\item LO FFs for $(p+\bar p)$:
\begin{description}
\item $D_u^{p/\bar p}(x,\mu^2)=2D_d^{p/\bar p}(x,\mu^2)$:
\begin{eqnarray}
N&=&0.40211-0.21633\bar s-0.07045\bar s^2+0.07831\bar s^3\nonumber\\
\alpha&=&-0.85973+0.13987\bar s-0.82412\bar s^2+0.43114\bar s^3\nonumber\\
\beta&=&2.80160+0.78923\bar s-0.05344\bar s^2+0.01460\bar s^3\nonumber\\
\gamma&=& 0.05198\bar s-0.04623\bar s^2
\end{eqnarray}
\item $D_s^{p/\bar p}(x,\mu^2)$:
\begin{eqnarray}
N&=&4.07885-2.97392\bar s-0.92973\bar s^2+1.23517\bar s^3\nonumber\\
\alpha&=&-0.09735+0.25834\bar s-1.52246\bar s^2+0.77060\bar s^3\nonumber\\
\beta&=&4.99191+1.14379\bar s-0.85320\bar s^2+0.45607\bar s^3\nonumber\\
\gamma&=&0.07174 \bar s-0.08321\bar s^2
\end{eqnarray}
\item $D_c^{p/\bar p}(x,\mu^2)$:
\begin{eqnarray}
N&=&0.11061-0.07726\bar s_c+0.05422\bar s_c^2-0.03364\bar s_c^3\nonumber\\
\alpha&=&-1.54340-0.20804\bar s_c+0.29038\bar s_c^2-0.23662\bar s_c^3
\nonumber\\
\beta&=&2.20681+0.62274\bar s_c+0.29713\bar s_c^2-0.21861\bar s_c^3\nonumber\\
\gamma&=&0.00831\bar s_c+0.00065\bar s_c^2
\end{eqnarray}
\item $D_b^{p/\bar p}(x,\mu^2)$:
\begin{eqnarray}
N&=&40.0971-123.531\bar s_b+128.666\bar s_b^2-29.1808\bar s_b^3\nonumber\\
\alpha&=&0.74249-1.29639\bar s_b-3.65003\bar s_b^2+3.05340\bar s_b^3
\nonumber\\
\beta&=&12.3729-1.04932\bar s_b+0.34662\bar s_b^2-1.34412\bar s_b^3\nonumber\\
\gamma&=&-0.04290\bar s_b-0.30359\bar s_b^2
\end{eqnarray}
\item $D_g^{p/\bar p}(x,\mu^2)$:
\begin{eqnarray}
N&=&0.73953-1.64519\bar s+1.01189\bar s^2-0.10175\bar s^3\nonumber\\
\alpha&=&-0.76986-3.58787\bar s+13.8025\bar s^2-13.8902\bar s^3\nonumber\\
\beta&=&7.69079-2.84470\bar s-0.36719\bar s^2-2.21825\bar s^3\nonumber\\
\gamma&=&1.26515\bar s-1.96117\bar s^2
\end{eqnarray}
\end{description}
\item NLO FFs for $(p+\bar p)$:
\begin{description}
\item $D_u^{p/\bar p}(x,\mu^2)=2D_d^{p/\bar p}(x,\mu^2)$:
\begin{eqnarray}
N&=&1.25946-1.17505\bar s+0.37550\bar s^2-0.01416\bar s^3\nonumber\\
\alpha&=&0.07124-0.29533\bar s-0.24540\bar s^2+0.16543\bar s^3\nonumber\\
\beta&=&4.12795+0.98867\bar s-0.46846\bar s^2+0.20750\bar s^3\nonumber\\
\gamma&=&0.18957\bar s-0.01116\bar s^2
\end{eqnarray}
\item $D_s^{p/\bar p}(x,\mu^2)$:
\begin{eqnarray}
N&=&4.01135+8.67124\bar s-22.7888\bar s^2+11.4720\bar s^3\nonumber\\
\alpha&=&0.17258+4.57608\bar s-9.64835\bar s^2+4.61792\bar s^3\nonumber\\
\beta&=&5.20766+7.25144\bar s-12.6313\bar s^2+6.07314\bar s^3\nonumber\\
\gamma&=&0.16931\bar s-0.09541\bar s^2 
\end{eqnarray}
\item $D_c^{p/\bar p}(x,\mu^2)$:
\begin{eqnarray}
N&=&0.08250-0.04512\bar s_c-0.00565\bar s_c^2+0.00900\bar s_c^3\nonumber\\
\alpha&=&-1.61290-0.38012\bar s_c-0.06840\bar s_c^2+0.08888\bar s_c^3
\nonumber\\
\beta&=&2.01255+0.63782\bar s_c-0.14146\bar s_c^2+0.06083\bar s_c^3\nonumber\\
\gamma&=&-0.02958\bar s_c+0.01130\bar s_c^2
\end{eqnarray}
\item $D_b^{p/\bar p}(x,\mu^2)$:
\begin{eqnarray}
N&=&24.2916-88.3524\bar s_b+93.1056\bar s_b^2-17.4089\bar s_b^3\nonumber\\
\alpha&=&0.57939-0.80783\bar s_b-5.07200\bar s_b^2-2.45377\bar s_b^3
\nonumber\\
\beta&=&12.1207-3.27370\bar s_b+1.21188\bar s_b^2-5.50374\bar s_b^3\nonumber\\
\gamma&=&0.14628\bar s_b-0.78634\bar s_b^2
\end{eqnarray}
\item $D_g^{p/\bar p}(x,\mu^2)$:
\begin{eqnarray}
N&=&1.56255-1.48158\bar s-0.39439\bar s^2+0.51249\bar s^3\nonumber\\
\alpha&=&0.01567-2.16232\bar s+2.47127\bar s^2-0.93259\bar s^3\nonumber\\
\beta&=&3.57583+3.33958\bar s-3.05265\bar s^2+1.21042\bar s^3\nonumber\\
\gamma&=&-0.84816\bar s+1.23583\bar s^2
\end{eqnarray}
\end{description}
\end{description}

\end{appendix}

\newpage
\begin{table}[ht]
\begin{center}
\caption{$\chi^2_{\mathrm{DF}}$ values and respective numbers of data points
used for their evaluation for the various data sets shown in
Figs.~\ref{fig:do}--\ref{fig:oy2}.
The selection criterion is $x>0.1$ in the case of $e^+e^-$ annihilation and
$p_T>\mu_0/\xi$ otherwise.
The cases where the experimental $p_T$ intervals contain values with
$p_T<\mu_0/\xi$ are marked by asterisks.}
\label{tab:one}
\medskip
\begin{tabular}{|c|r|c|rr|rr|rr|} \hline\hline
Type & Energy & Experiment & \multicolumn{6}{c|}{$\xi$} \\
\cline{4-9}
 & [GeV] & & \multicolumn{2}{c|}{1/2} & \multicolumn{2}{c|}{1} &
\multicolumn{2}{c|}{2} \\
\hline
$e^+e^-$ & 133 & DELPHI \cite{D2} & 0.514 & 6 & 0.322 & 6 & 0.498 & 6 \\
         &     & OPAL   \cite{O2} & 0.413 & 11 & 0.317 & 11 & 0.404 & 11 \\
         & 161 & DELPHI \cite{D2} & 0.608 & 6 & 0.344 & 6 & 0.360 & 6 \\
         &     & OPAL   \cite{O2} & 0.532 & 11 & 0.353 & 11 & 0.361 & 11 \\
         & 172 & DELPHI \cite{D2} & 1.01 & 6 & 1.10 & 6 & 1.26 & 6 \\
         &     & OPAL   \cite{O2} & 0.902 & 11 & 1.04 & 11 & 1.25 & 11 \\
         & 183 & DELPHI \cite{D2} & 0.498 & 6 & 0.599 & 6 & 0.877 & 6 \\
         &     & OPAL   \cite{O2} & 1.20 & 11 & 1.64 & 11 & 2.38 & 11 \\
         & 189 & OPAL   \cite{O2} & 0.191 & 11 & 0.568 & 11 & 1.57 & 11 \\
\hline
$p\bar p$ &  200 & UA1 \cite{UA1b} & 0.976 & 14 & 17.9 & 28 & 48.4 & 35 \\
          &  500 & UA1 \cite{UA1b} & 6.43 & 14 & 5.13 & 28 & 22.6 & 35 \\
          &  540 & UA1 \cite{UA1a} & 10.4 & 14 & 3.19 & 28 & 18.9 & 35 \\
          &      & UA2 \cite{UA2a} & -- & -- & 2.30 & 6 & 60.0 & 13 \\
          &      & UA2 \cite{UA2b} & 1.36 & 12 & 3.99 & 22 & 21.0 & 29 \\
          &  630 & UA1 \cite{UA1c} & 7.68 & 33 & 1880 & 40 & 8360 & 43 \\
          &      & CDF \cite{CDF}  & 14.3 & 2 & 2.04 & 12 & 47.3 & 22 \\
          &  900 & UA1 \cite{UA1b} & 27.0 & 26 & 7.49 & 40 & 27.2 & 47 \\
          & 1800 & CDF \cite{CDF}  & 38.5 & 19 & 6.56 & 33 & 433 & 43 \\
\hline
$\gamma p$ & 165--251 & H1 \cite{H1}    & 9.89 & 18 & 0.467 & 26 & 4.15 & 26 \\
           &          & $p_T>2$~GeV  & 59.9$^*$ & 10 & 1.01 & 10 & 8.11 & 10 \\
           &          & $p_T>3$~GeV     & 14.8 & 10 & 0.564 & 10 & 4.11 & 10 \\
           & 167--194 & ZEUS \cite{ZEUS} & 12.7 & 11 & 3.79 & 23 & 2.06 & 30 \\
\hline
$\gamma\gamma$ & 10--30 & OPAL \cite{Ogg} & 2.58 & 5 & 5.38 & 12 & 19.5 & 15 \\
               &        & $p_T>1.5$~GeV & 6.61$^*$ & 5 & 15.3 & 5 & 24.6 & 5 \\
               & 30--55 & OPAL \cite{Ogg} & 1.06 & 7 & 1.70 & 14 & 6.65 & 17 \\
               &        & $p_T>1.5$~GeV & 97.8$^*$ & 5 & 1.74 & 5 & 12.4 & 5 \\
               & 55--125 & OPAL \cite{Ogg} & 2.13 & 7 & 2.33 & 14 & 2.95 & 17 \\
               &        & $p_T>1.5$~GeV & 235$^*$ & 5 & 3.56 & 5 & 6.89 & 5 \\
               & 10--125 & OPAL \cite{Ogg} & 1.73 & 7 & 3.25 & 14 & 14.4 & 17 \\
               &        & $p_T>1.5$~GeV & 70.3$^*$ & 5 & 6.61 & 5 & 26.1 & 5 \\
\hline\hline
\end{tabular}
\end{center}
\end{table}

\newpage
\begin{table}[ht]
\begin{center}
\caption{$\chi^2_{\mathrm{DF}}$ values and respective numbers of data points
used for their evaluation for the various $p_T$ distributions shown in
Figs.~\ref{fig:u1a}--\ref{fig:hp}, \ref{fig:z}, and \ref{fig:op1}.
The selection criterion is $p_T>2\mu_0$.}
\label{tab:two}
\medskip
\begin{tabular}{|c|r|c|r|r|r|r|} \hline\hline
Type & Energy & Experiment & \multicolumn{3}{c|}{$\xi$} & No.\ of \\
\cline{4-6}
 & [GeV] & & \multicolumn{1}{c|}{1/2} & \multicolumn{1}{c|}{1} &
\multicolumn{1}{c|}{2} & Points \\
\hline
$p\bar p$ &  200 & UA1 \cite{UA1b} & 0.976 & 4.93 & 9.68 & 14 \\
          &  500 & UA1 \cite{UA1b} & 6.43 & 1.53 & 4.20 & 14 \\
          &  540 & UA1 \cite{UA1a} & 10.4 & 1.44 & 3.32 & 14 \\
          &      & UA2 \cite{UA2b} & 1.36 & 3.83 & 6.94 & 12 \\
          &  630 & UA1 \cite{UA1c} & 7.68 & 354 & 756 & 33 \\
          &      & CDF \cite{CDF}  & 14.3 & 1.17 & 2.07 & 2 \\
          &  900 & UA1 \cite{UA1b} & 27.0 & 4.08 & 9.41 & 26 \\
          & 1800 & CDF \cite{CDF}  & 38.5 & 2.52 & 5.59 & 19 \\
\hline
$\gamma p$ & 165--251 & H1 \cite{H1}     & 9.89 & 0.416 & 2.65 & 18 \\
           & 167--194 & ZEUS \cite{ZEUS} & 12.7 & 3.35 & 0.443 & 11 \\
\hline
$\gamma\gamma$ &  10--30 & OPAL \cite{Ogg} & 2.58 & 2.18 & 2.80 & 5 \\
               &  30--55 & OPAL \cite{Ogg} & 1.06 & 1.31 & 1.70 & 7 \\
               & 55--125 & OPAL \cite{Ogg} & 2.13 & 0.110 & 0.467 & 7 \\
               & 10--125 & OPAL \cite{Ogg} & 1.73 & 2.44 & 4.31 & 7 \\
\hline\hline
\end{tabular}
\end{center}
\end{table}

\newpage
\begin{table}[ht]
\begin{center}
\caption{$\chi^2_{\mathrm{DF}}$ values and respective numbers of data points
used for their evaluation for the various $x$ distributions shown in
Figs.~\ref{fig:c1}--\ref{fig:c3}.
The selection criterion is $x>0.1$.
Data samples which did not enter the respective fits are marked by asterisks.}
\label{tab:three}
\medskip
\begin{tabular}{|r|c|c|r|r|r|r|} \hline\hline
Energy & Flavour & Experiment & \multicolumn{3}{c|}{FF Set} & No.\ of \\
\cline{4-6}
[GeV] & & & KKP \cite{kkp} & K \cite{kre} & BFGW \cite{bou} & Points \\
\hline
29 & $uds$ & TPC \cite{T} & $0.178^*$ & 0.159 & $0.167^*$ & 7 \\
   & $c$   &              & $0.876^*$ & 0.911 & $0.923^*$ & 7 \\
   & $b$   &              & $2.23^*$ & 1.21 & $1.14^*$ & 7 \\
\hline
91.2 & all   & DELPHI \cite{D1} & 1.28 & $1.51^*$ & 1.49 & 12 \\
     &       & SLD \cite{S}     & 1.32 & 0.370 & 0.421 & 21 \\
     & $uds$ & DELPHI \cite{D}  & $3.17^*$ & $0.990^*$ & 1.95 & 13 \\
     &       & DELPHI \cite{D1} & 0.201 & $0.588^*$ & $1.00^*$ & 12 \\
     & $c$   & DELPHI \cite{D}  & $0.473^*$ & $0.388^*$ & 0.401 & 11 \\
     & $b$   & DELPHI \cite{D}  & $28.9^*$ & $0.887^*$ & 1.03 & 12 \\
     &       & DELPHI \cite{D1} & 0.433 & $9.14^*$ & 8.74 & 12 \\
\hline
189  & all & OPAL \cite{O2} & $0.568^*$ & $0.250^*$ & $0.414^*$ & 11 \\
\hline\hline
\end{tabular}
\end{center}
\end{table}

\newpage

\newpage

\newpage

\newpage

\newpage

\newpage

\newpage

\newpage

\newpage

\newpage

\newpage

\newpage

\newpage

\newpage

\newpage

\end{document}